%% file: LiveSpeechPortraits.tex
\documentclass[acmtog]{acmart}
\acmSubmissionID{145}

\usepackage{booktabs} 

\usepackage{hyperref} 

\usepackage[ruled]{algorithm2e} 

\usepackage{xcolor}
\SetAlFnt{\small}
\SetAlCapFnt{\small}
\SetAlCapNameFnt{\small}
\SetAlCapHSkip{0pt}

\usepackage{float}
\usepackage{dblfloatfix}
\usepackage{multirow}

\setcopyright{acmcopyright}
\acmJournal{TOG}
\acmYear{2021}
\acmVolume{40}
\acmNumber{6}
\acmArticle{220}
\acmMonth{12} 
\acmDOI{10.1145/3478513.3480484}

\begin{document}
\title{Live Speech Portraits: Real-Time Photorealistic Talking-Head Animation}

\author{Yuanxun Lu}
\affiliation{%
  \institution{Nanjing University and Xmov}
  \country{China}}
\email{luyuanxun@smail.nju.edu.cn}
\author{Jinxiang Chai}
\authornote{Corresponding authors}
\affiliation{%
  \institution{Xmov}
  \city{Shanghai}
  \country{China}}
\email{chaijinxiang@xmov.ai}
\author{Xun Cao}
\authornotemark[1]
\affiliation{%
  \institution{Nanjing University}
  \city{Nanjing}
  \country{China}}
\email{caoxun@nju.edu.cn}

\begin{abstract}
To the best of our knowledge, we first present a live system that generates personalized photorealistic talking-head animation only driven by audio signals at over 30 fps. Our system contains three stages. The first stage is a deep neural network that extracts deep audio features along with a manifold projection to project the features to the target person's speech space. In the second stage, we learn facial dynamics and motions from the projected audio features. The predicted motions include head poses and upper body motions, where the former is generated by an autoregressive probabilistic model which models the head pose distribution of the target person. Upper body motions are deduced from head poses. In the final stage, we generate conditional feature maps from previous predictions and send them with a candidate image set to an image-to-image translation network to synthesize photorealistic renderings. Our method generalizes well to wild audio and successfully synthesizes high-fidelity personalized facial details, e.g., wrinkles, teeth. Our method also allows explicit control of head poses. Extensive qualitative and quantitative evaluations, along with user studies, demonstrate the superiority of our method over state-of-the-art techniques. 

\footnotetext{This project was performed during Yuanxun Lu's internship at Xmov. Project website: \href{https://yuanxunlu.github.io/projects/LiveSpeechPortraits/}{https://yuanxunlu.github.io/projects/LiveSpeechPortraits/}}.

\end{abstract}

%
%



%
%


\begin{teaserfigure}
  \includegraphics[width=\textwidth, scale=0.5]{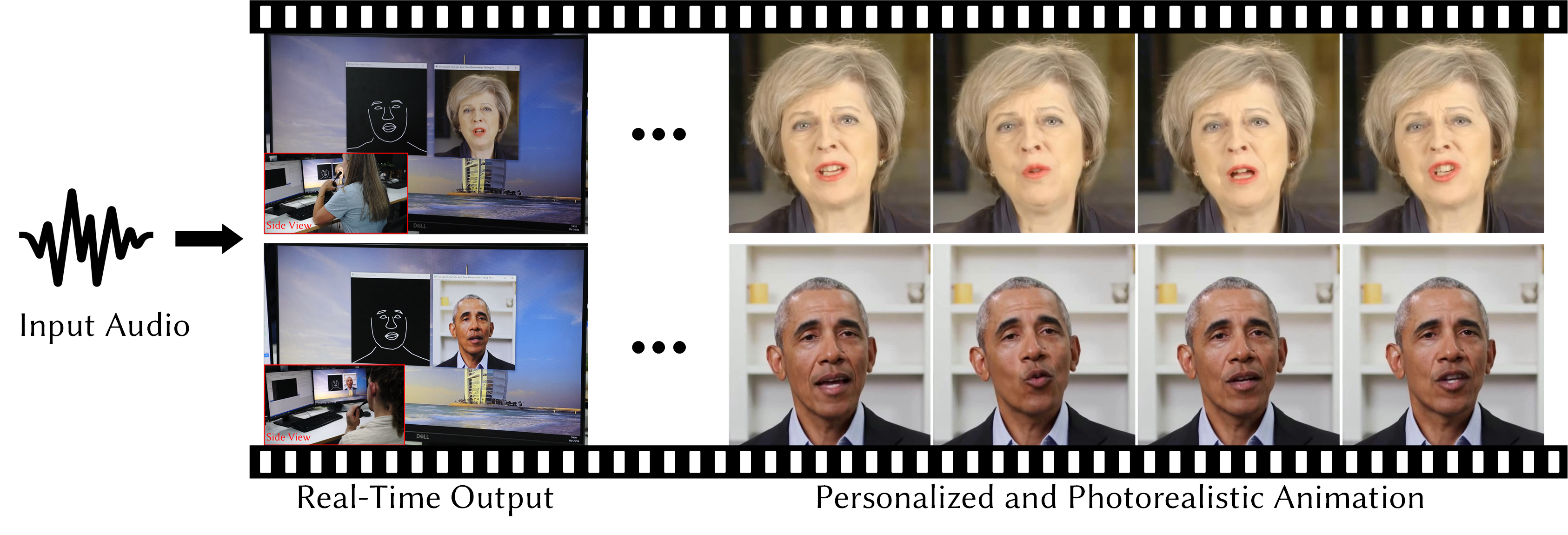}
  \caption{Given an arbitrary input audio stream, our system generates personalized and photorealistic talking-head animation in real-time. Right: May and Obama are driven by the same utterance but present different speaking characteristics. Video \emph{May} \copyright \emph{UK government} (Open Government Licence). Video \emph{Obama} \copyright \emph{Barack Obama Foundation} (public domain).}
  \label{fig:teaser}
\end{teaserfigure}

\maketitle

\input{Source-LiveSpeechPortraits}

\end{document}

%% file: Source-LiveSpeechPortraits.tex
\section{Introduction}

Talking-head animation, i.e., synthesizing audio-synchronized video frames of a target person, is valuable to interactive applications like digital avatars, video conferencing, visual effects, virtual reality, visual dubbing and computer games. With recent advances in deep learning, people have made great progress in this long-standing problem. However, achieving a realistic and expressive talking-head animation remains an open challenge. Humans are extremely sensitive to any facial artifacts, leading to high requirements for the desired techniques.

Several factors contribute to the challenge. Firstly, attempts to generate lip-synchronized and personalized facial dynamics face a two-fold difficulty, due in part to the challenge of mapping from 1-D audio signals to facial movements which lie on high-dimensional manifolds, but also due to the domain difference between wild audio and target speech space, which makes the system fail to preserve individual talking idiosyncrasies. Secondly, head and body motion, another critical component of lifelike animation, is not closely correlated to audio. For example, one can swing his head or be still when he says the same words, which depends on many factors - his mood, location, or history poses. Thirdly, synthesizing controllable photorealistic renderings of the target is non-trivial. Nowadays, traditional rendering engines are still far from desired, whose results can be recognized as fake at a glance. Neural renderer shows great power on photorealistic rendering but suffers from performance degradation if the predicted motion is far outside the span of the training corpus \cite{kim2018deep}. Last but not least, many interactive scenarios like video conferencing and digital avatars require the entire system to run in real-time, which makes high demands of the system efficiency without damaging the performance.

In this paper, we propose a deep learning architecture, called Live Speech Portraits (LSP), to address these challenges and step further to practical applications. Our system generates personalized talking-head animation stream, including facial expressions and motion dynamics (head pose and upper body motion) driven by audio and allows for photorealistic rendering in real-time. 

First of all, we adopt the idea of self-supervised representation learning, which has shown great power in learning semantic or structural representations and benefits various downstream tasks \cite{he2020momentum, chen2020simple, CPC}, to extract speaker-independent audio features. To achieve realistic and personalized animation on wild audio streams, we further project the wild features to the target feature space and reconstruct them using target features. This process can be seen as domain adaption from source to target. Subsequently, we are able to learn the mapping from reconstructed audio features to facial dynamics.

Another critical component that contributes to realistic talking-head animation is head and body motion. To generate personalized and time-coherent head poses from audio, we make assumptions that the current head pose is \emph{partly} correlated to audio information and also \emph{partly} to history poses. We propose a novel autoregressive probabilistic model to learn the head pose distribution of the target person based on these two conditions. Head poses are sampled from the estimated distribution, and upper body motions are futher deduced from the sampled head poses.

To synthesize photorealistic renderings, we employ an image-to-image translation network conditioned on a feature map and candidate images. We apply the sampled rigid head poses on facial dynamics and project the transformed facial keypoints and upper body positions to the image plane, generating landmark images as our intermediate representations. Although our system comprises several modules, it is still compact enough to run in real-time at over 30 fps. In summary, we present the following contributions:

\begin{itemize}
\item To the best of our knowledge, we propose Live Speech Portraits (LSP) as the first audio-driven talking-head animation system with photorealistic renderings in real-time. A comprehensive evaluation demonstrates that our approach outperforms prior methods both qualitatively and quantitatively. 

\item A novel audio feature extraction module that generalizes our system to wild audio signals. The key component of this module is a manifold projection that reconstructs the deep speech representations using target speech features.

\item An elaborately designed probabilistic autoregressive architecture that predicts personalized head pose distributions conditioned on audio signals and history motions. Our system also allows for user-controllable head poses generation.

\end{itemize}

\begin{figure*}[h]
    \centering
  \includegraphics[width=\linewidth, scale=0.5]{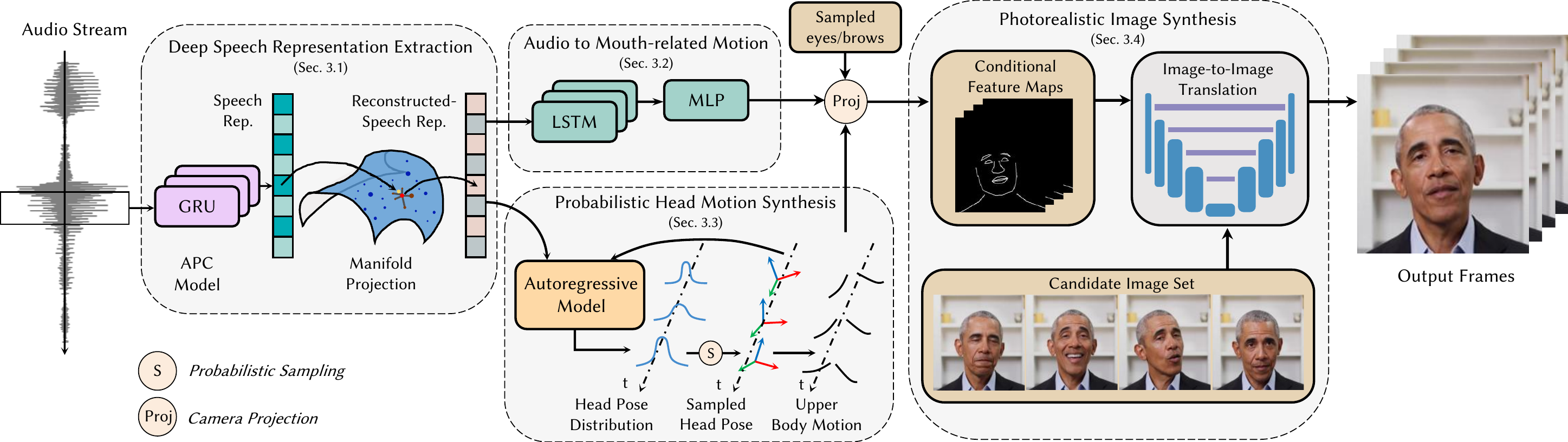}
  \caption{An overview of our \emph{Live Speech Portraits} method. Given an arbitrary audio stream, our method generates personalized and photorealistic talking animation of a target person in real-time. First, deep speech representations of the input audio are extracted and reconstructed using manifold projection. Then, mouth-related motions, head poses, and upper body motions are predicted from the reconstructed speech representations. We then generate conditional feature maps by projecting the predicted motions and other sampled facial components. Finally, we send conditional feature maps and a candidate image set to an image-to-image translation network to synthesize photorealistic talking portraits. Video \emph{Obama} \copyright \emph{Barack Obama Foundation} (public domain).}
  \label{fig:pipeline}
\end{figure*}

\section{Related Work}
\label{section 2}
Mathematically, audio-driven facial animation aims to generate a sequence of talking-head frames from an input audio stream. In the following, we generally review prior work on audio-driven facial animation, as well as related techniques on speech representation learning, head pose estimation and facial reenactment.\\

\emph{Audio-driven Talking-head Animation.} Audio-driven talking-head animation is a cross-modal research topic with a long history in the computer graphics community. Prior approaches take two different roads depending on whether they aim for generating photorealistic videos. In the non-photorealistic case, these methods focus on learning a mapping from input waveforms to facial movements, e.g., 3D vertex coordinates \cite{SIG17NVIDIA, VOCA}, reference facial model parameters \cite{SIG17CMU} or rigging parameters \cite{visemenet}. These methods usually require high-quality 4D face capture data or rigging parameters with artist interventions. Here, we focus on the photorealistic case which our method belongs to. More than twenty years ago, people have made groundbreaking explorations in this field. Bregler et al. \shortcite{VideoRewrite} proposed \emph{Video Rewrite} to create a new person-talking video using existing footage. Brand \shortcite{voicepuppetry} proposed \emph{Voice Puppetry} to generate full facial animation from an audio track. These techniques can roughly be categorized into video-based editing methods and image-based generation methods. Video-based editing methods yield editing on a target video - usually synthesize a mouth-related region patch and blend it into the target frame while keeping other regions unaltered \cite{ezzat2002trainable, garrido2015vdub, NVP}. Recently, Thies et al. \shortcite{NVP} proposed \emph{Neural Voice Puppetry} as an upgrade to \emph{Voice Puppetry}. They first learned a generalized 3D face model from audio sequences and then fine-tuned the model on the target clip via learning a person-specific blendshape basis, in which case the talking style of the target portrait can be preserved. Lower faces were synthesized via a neural rendering network finally. However, these approaches have several intrinsic limitations. First, animation length is restricted to target video length. A heuristic post-processing to select proper candidate frames is required for generating longer videos. Second, head poses and upper body motions are uncontrollable since the are directly copied from target videos, which may conflict with the audio track and introduce barriers to real-time applications. Notably, Suwajanakorn et al. \shortcite{SIG17Obama} employed a re-timing schedule to select target frames with natural and synchronized head motions. Last but not least, these methods rely on successful face tracking and tend to fail when faces are partly unseen or undetected, e.g., lower faces are obscured by hands or in a very dark environment. Skipping these bad frames leads to a temporally inconsistent result. In contrast, our method synthesizes portraits directly. Obstructed frames could be dropped before training without affecting performance.

Image-based generation methods generate a talking-head video based on one or several cropped reference images. This kind of method avoids the pre-mentioned drawbacks but makes the task more challenging for the requirements of manipulating the entire image, including facial details, motion dynamics, and the background. End-to-end training \cite{chung2017you, wiles2018x2face, zhou2021pose} is becoming a strong trend to generate videos with the rising of deep learning. Chuang et al. \shortcite{chung2017you} generated a talking face video from a still image and an audio sequence by utilizing a CNN model for the first time. Later, GANs are frequently adopted to generate high-fidelity facial images via adversarial learning \cite{vougioukas2018end, vougioukas2019realistic, zhou2019talking}. Instead of directly synthesizing a talking face image, Chen et al. \shortcite{ATVG} and Zhou et al. \shortcite{makelttalk} leveraged sparse facial landmarks as an intermediate representation. The landmark dynamics were first deduced from the audio input through an audio-to-landmark module and then worked as a condition of an image-to-image translation network to generate animated videos. One common problem shared by these methods is that they tend to learn average facial dynamics among the training corpus without \emph{person-specific} talking styles. Note that Zhou et al. \shortcite{makelttalk} learned \emph{speaker-aware} dynamics from speaker embedding vectors, but still fail to learn \emph{target-aware} dynamics which may generate uncanny results. Our method focuses on capturing the \emph{person-specific talking dynamics} using only a short target video (around 3 minutes). We utilize facial landmarks as an intermediate representation and generate controllable head poses and upper body motions, which makes the animated videos more impressive and realistic.\\

\emph{Speech Representation Learning}. Speech signals contain rich high-level information, including content, timbre, and prosody. Much prior work require accurate phoneme labels within millisecond timestamps as input. These labels are often assembled into a sequence of diphones or triphones to encode neighborhood information \cite{fan2015photo}. However, converting waveforms to phonemes leads to information compression, along with potential performance reduction introduced by error-prone automatic phoneme labeling tools. People also discovered different schemes to get rid of the dependence of phonemes using hand-crafted features \cite{SIG17Obama}. Recently, modeling these semantic and structural representations through deep neural networks has shown great success and outperforms the traditional hand-crafted features \cite{devlin2018bert, peters2018deep}. Thies et al. \shortcite{NVP} employed a DeepSpeech \cite{hannun2014deep} network to extract speech features. Zhou et al. \shortcite{makelttalk} resorted to voice conversion community \cite{autovc} to disentangle speech content and identity information. Similarly, Our system uses a self-supervised learning method \cite{APC} to extract the high-level speech information. Moreover, manifold projection is applied to improve generalization.\\

\emph{Head Pose Estimation from Audio}. Head pose, as a significant component of realistic animation, delivers rich information in talking-head videos. Greenwood et al. \shortcite{greenwood2018predicting} employed a Bi-directional LSTM model to predict character head animation from audio. Zhou et al. \shortcite{makelttalk} predicted speaker-aware head motion dynamics as 3D facial landmark displacements. They trained a transformer architecture \cite{transformer} in an adversarial mechanism to capture long time dependencies and generate natural head dynamics. Recently, Chen et al. \shortcite{chen2020talking} proposed a 3D-aware generative network to learn target-aware head motion from a 3-second video clip. Different from most previous work which use deterministic models, we use an auto-regressive probabilistic model conditioned on history head poses and speech representations to predict the distribution parameters at the current timestamp. Head poses are sampled from the predicted probabilistic model. Besides, we further deduce upper body motions from the predicted head pose, which shows a great improvement on the animation quality.\\

\emph{Video-based Facial Reenactment.} Video-based facial reenactment is another technique related to audio-driven animation. Thies et al. \shortcite{thies2015realtime} proposed the first real-time model-based reenactment system using two RGBD cameras. Face2Face \cite{thies2016face2face} extended the boundary using only RGB cameras. Moreover, Liu et al. \shortcite{liu2015video} combined both audio and visual information as input and tackles the problem that tracking results prone to fail when face is occluded or head pose is extreme. Fried et al. \shortcite{SIG19-text} proposed a method for text-based talking-head editing while is slow for viseme search (5 mins for three words). Yao et al. \shortcite{yao2021iterative} reduced the video generation time to 40 seconds for one video. Recently, GANs achieved great success in controllable high-fidelity face synthesis \cite{karras2019style, wang2018high, wang2018video}. Few shot or even one-shot facial animation methods are explored via landmarks pre-defined or learned in an unsupervised scheme \cite{zakharov2019few, siarohin2019first, sun2020landmarkgan}. Most methods rely on an image-to-image mechanism with semantic images as input. Kim et al. \shortcite{kim2018deep} generated portrait videos including head, mouth and gaze from an input reference video. Kim et al. \shortcite{kim2019neural} trained a recurrent GAN to synthesize style-preserving visual dubbing. Very recently, Elgharib at al. \shortcite{elgharib2020egocentric} transferred the egocentric view videos to front facing videos using a position conditional GAN. Different from previous methods, our approach generates photorealistic talking head animation from speech only and runs in real-time.

\section{Method}
\emph{Overview.} Given an arbitrary speech stream, our live speech portraits approach generates photorealistic talking-head animation of the \emph{target person} in real-time (Figure \ref{fig:pipeline}). Our approach consists of three stages: deep speech representation extraction, audio-to-face prediction, and photorealistic face rendering. The first stage extracts the speech representation of the input audio (Section \ref{section 3.1}). The representation extractor learns the high-level speech representation and is trained in a self-supervised manner on an unlabelled speech corpus. We then project the representations to the target person's speech space to improve generalization. The second stage predicts the full motion dynamics. Two elaborate designed neural networks predict the mouth-related motion (Section \ref{section 3.2}) and head pose (Section \ref{section 3.3}) from the speech representations, respectively. The mouth-related motions are represented as sparse 3D landmarks, and head poses are represented as rigid rotation and translation. Considering that head poses are less related to audio information than mouth-related motions, we employ a probabilistic autoregressive model to learn the poses conditioned on audio information and history poses. Other facial components which have nearly no correlation to audio (e.g., eyes, brows, noses, etc.) are sampled from the training set. We then compute the upper body motion from the predicted head pose. The final stage synthesizes the photorealistic video frames from the previous predictions and a candidate image set (Section \ref{section 3.4}) using a conditional image-to-image translation network. In the following, we introduce each module in detail.

\subsection{Deep Speech Representation Extraction}
\label{section 3.1}
Input information, which are speech signals in our case, plays a crucial role because it powers the entire system. As illustrated in Section \ref{section 2}, people have exploited deep learning approaches, commonly trained in a self-supervised mechanism, to learn high-level speaker-independent representations of speech from surface features. These methods greatly improve state-of-the-art performance of downstream tasks, e.g., auto speech recognition, phone classification, and speaker verification \cite{CPC, chorowski2019unsupervised, NPC}. 

Specifically, we use the autoregressive predictive coding (APC) model \cite{APC} to extract structural speech representations. The APC model predicts future surface features given history information. In our case, we select 80-dimensional log Mel spectrograms as speech surface features. The model is a standard 3-layer unidirectional Gated Recurrent Units (GRUs):
\begin{equation}
    \textbf{h}_l = GRU^{(l)}(\textbf{h}_{l-1}), \forall l \in [1, L],
\end{equation}
where $\textbf{h}_l \in \mathbb{R}^{512} $ is the hidden states of each layer in GRUs. The hidden states in the final GRU layer are our desired deep speech representations. We add a linear layer to map the output to predict the future log Mel spectrograms during training, and the linear layer is dropped during testing time.

\begin{figure}[t]
  \centering
  \includegraphics[width=\linewidth]{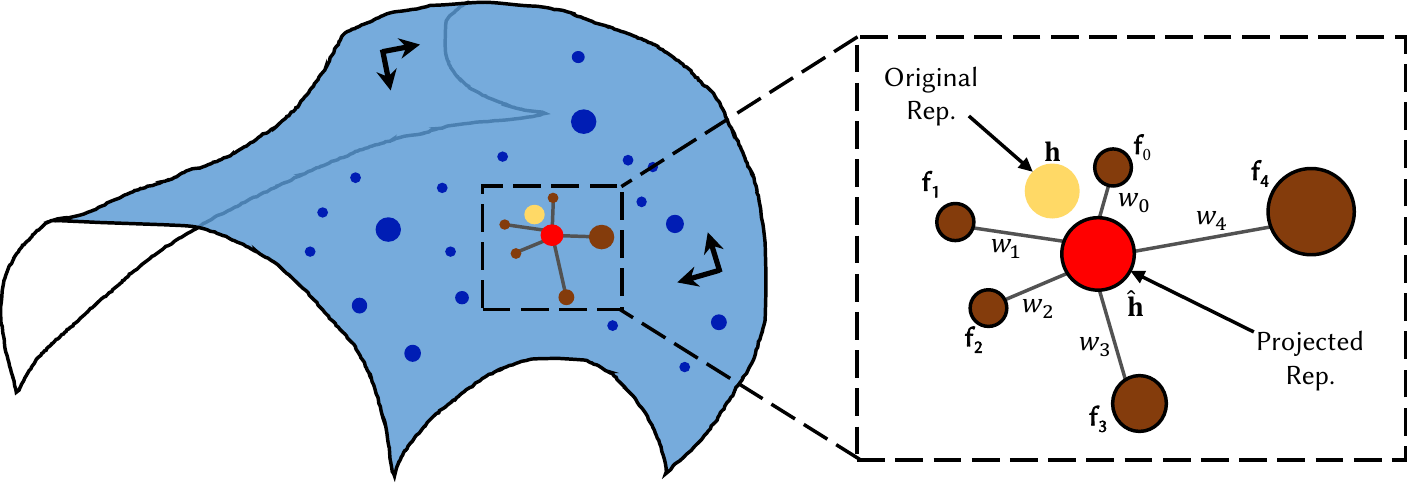}
  \caption{Manifold projection. Left: For each original deep feature, we project it to the target feature space. Right: Zoom in of the original feature (yellow), $k=5$ nearest neighbours (brown), and the reconstructed feature (red).}
  \label{fig:manifold_projection}
\end{figure}

\subsubsection{Manifold Projection} 
\label{section 3.1.1}
Different person owns diverse speaking styles, which are considered as personalized styles. For example, \emph{May clip} exhibits large lip movements and always the 'O' style, \emph{Ford clip} exhibits small lip movements like whispers, and \emph{Nadella clip} exhibits adhesion of upper \& lower lips like lisps. Directly applying the deep speech representation may lead to poor results when input speech representations locate far away from the target's speech feature space (e.g., animate a woman by men's voice, foreign languages, or even songs). We perform manifold projection after extracting the speech representations to improve generalization.

The manifold projection operation is inspired by the recent success in face synthesis from sketches \cite{chen2020deepfacedrawing}, which can be generalized to sketches far away from human faces. We apply the locally linear embedding (LLE) assumption on the speech representation manifold: each data point and its neighbors are locally-linear on the high-dimensional manifold \cite{roweis2000nonlinear}.

Given an extracted speech representation $\textbf{h} \in \mathbb{R}^{512}$, we compute the LLE reconstructed representation $\hat{\textbf{h}} \in \mathbb{R}^{512}$ on each dimension. As illustrated in Figure \ref{fig:manifold_projection}, we first find the $K$ nearest points of $\textbf{h}$ in a target speech representation database $\mathcal{D} \in \mathbb{R}^{N_s \times 512}$ by computing the Euclidean distance. $N_s$ is the number of training frames. Then, we seek a linear combination of the $K$ nearest neighbours to best reconstruct $\textbf{h}$, which is equivalent to compute the barycentric coordinates of $\textbf{h}$ based on its neighbors via solving the following minimization problem:
\begin{equation}
    \min || \textbf{h} - \sum_{k=1}^{K} w_k \cdot \textbf{f}_k ||_2^2, \quad s.t. \sum_{k=1}^{K} w_k = 1,
\end{equation}
where $w_k$ is the barycentric weight of $k$-nearest neighbor $\textbf{f}_k$, which can be computed via solving a least-squares problem. $K$ is chosen as 10 empirically in our experiment. At last, we obtain the projected speech representation $\hat{\textbf{h}}$:
\begin{equation}
    \hat{\textbf{h}} = \sum_{k=1}^{K} w_k \cdot \textbf{f}_k.
\end{equation}
Subsequently, $\hat{\textbf{h}}$ is sent to the motion predictors in Section \ref{section 3.2} and Section \ref{section 3.3} as input deep speech representations. Our experiment results show that the manifold projection helps improve our system generalization ability. Non-linear projection hasn't been considered for the complexity.

\subsection{Audio to Mouth-related Motion}
\label{section 3.2}
Predicting mouth-related motion from audio has been widely researched in the past few years. People use deep learning architectures to learn a mapping from audio features to intermediate representations, e.g., lip-related landmarks \cite{greenwood2018joint, makelttalk}, parameters of a parametric model \cite{SIG17Obama, SIG17CMU, ATVG}, 3D vertices \cite{SIG17NVIDIA, VOCA}, or facial blendshapes \cite{NVP, visemenet}. In our case, we use 3D displacements $\Delta \textbf{v}_m \in \mathbb{R}^{25 \times 3}$ with respect to mean positions of the target person in object coordinates as our intermediate representations.

To model sequence dependencies, we use Long Short Term Memory (LSTM) models to learn the mapping from speech representations to mouth-related motions. Similar to \cite{SIG17Obama}, we add a $d$ frames delay to make the model accessible to a short future, leading to significant improvement on the quality. We later feed the output of the LSTM network to a Multi-Layer Perception (MLP) and finally predict the 3D displacements $\Delta \textbf{v}_m$. In summary, our mouth-related prediction module works as follows:
\begin{flalign}
    \textbf{m}_0,\textbf{m}_1, ..., \textbf{m}_t &= LSTM(\hat{\textbf{h}}_0, \hat{\textbf{h}}_1, ..., \hat{\textbf{h}}_{t+d}), \\
    \Delta \textbf{v}_{m,t} &= MLP(\textbf{m}_t),
\end{flalign}
where time delay $d$ is set to 18 frames, equal to 300 ms delay in our experiments (60 FPS). The LSTM is stacked into three layers, and each layer has a hidden state of size 256. The MLP decoder network has three layers with hidden state size of 256, 512, and 75.

\begin{figure}[h]
  \centering
  \includegraphics[scale = 0.3]{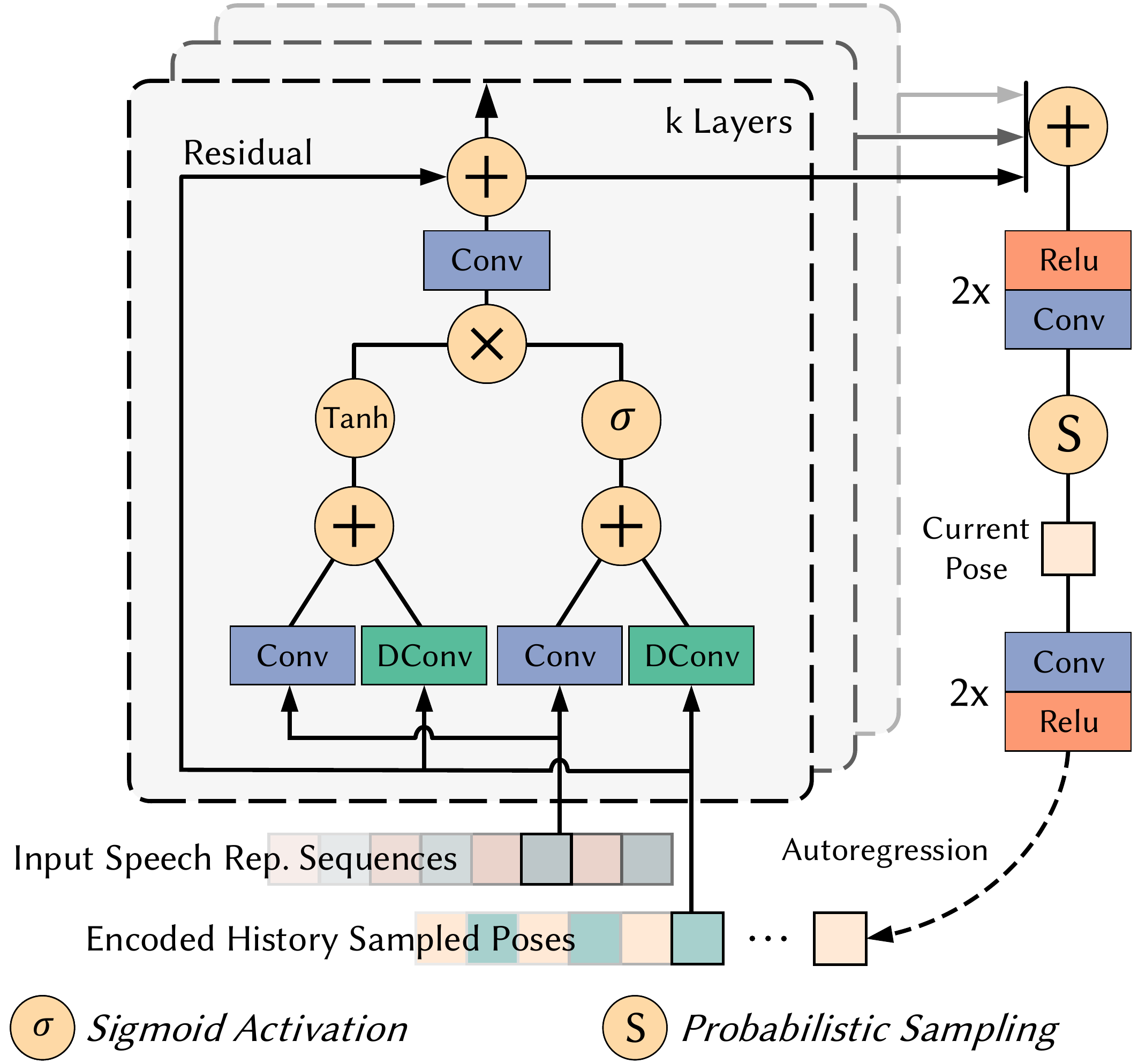}
  \caption{Illustration of our probabilistic head pose estimation network. This figure demonstrates an example architecture composed of one residual block with three layers.}
  \label{fig:wavenet}
\end{figure}

\subsection{Probabilistic Head and Upper Body Motion Synthesis}
\label{section 3.3}
Head pose and upper body motion are another two components that contribute to vivid talking-head animation. For example, people naturally swing their heads and move their bodies when talking, aiming to express emotion and deliver attitudes to the audience. We first describe the method to estimate head pose and then the upper body motion.

Head pose estimation from audio is non-trivial since little relationship exists between them. Considering the intrinsic difficulties, which is a one-to-many mapping from audio to head pose (one can say the same sentence in arbitrary poses), we make two assumptions as prior knowledge.\\

\textsc{Assumption} 1. \emph{Head poses are partly related to audio information, like expression and intonation. For example, people tend to nod heads when expressing agreement and look up when speaking in rising intonation and vice versa.}\\

\textsc{Assumption} 2. \emph{Current head pose partly depends on history head poses. For example, there is a large probability that people will turn heads back if they have turned a large angle before.}\\

These two assumptions simplify the problem and motivate our architecture design. To satisfy the requirements, the proposed network  $\Phi$ should have the ability to see the history head poses and current audio information as conditions. Besides, instead of considering it as a regression problem and train it using Euclidean distance loss \cite{makelttalk}, we should model this mapping as a probabilistic distribution. Recently, probabilistic models are successfully used in motion synthesis \cite{henter2020moglow} and outperform the deterministic models. The joint probability of the head motion can be described as follows:
\begin{flalign}
    p(\textbf{x} | \hat{\textbf{h}}) = \prod_{t=1}^T p(x_t | x_1, x_2, ..., x_{t-1}, \hat{\textbf{h}}_t)
\end{flalign}
where $x$ is the head motion and $\hat{\textbf{h}}$ is the speech representation.

The probabilistic model we use is a multi-dimensional Gaussian distribution. The network architecture is inspired by recent success in conditional probabilistic generative modeling \cite{oord2016conditional, wavenet}. The detailed design of the probabilistic model is illustrated in Figure \ref{fig:wavenet}. The model is a stack of two residual blocks with seven layers each. Considering the long-time dependencies required to produce natural head motion (one swing his head from left to right may lasts for several seconds), these residual blocks use dilation convolutions layers to capture the dependencies instead of normal convolutions with much fewer parameters. The dilation is doubled seven times for each layer in our architecture and then repeated twice: 1, 2, 4, 8, 16, 32, 64, 1, 2, 4, 8, 16, 32, 64. As a result, the history receptive field size $F$ of our model is 255 frames, equal to 4.25 seconds in our experiments. The output of each layer is summed up and processed by a post-processing network (a stack of 2 relu-conv layers) to generate the current distribution. In particular, the model outputs the mean values $\mu$ and the standard deviations $\sigma$ of the estimated Gaussian. Then we sample in the distribution to get final rigid head pose $P \in \mathbb{R}^{6}$, composed of 3D rotation $R \in \mathbb{R}^3$ and translation $T \in \mathbb{R}^3$.
We also tried with a Gaussian Mixture Model but found no obvious improvement. After sampling, we encode the current pose as input pose information for the next time step, forming an autoregressive mechanism. In summary, the head pose estimation can be illustrated as follows:
\begin{flalign}
    \textbf{P}_{para, t} &= \Phi(\textbf{P}_{t-F}, ..., \textbf{P}_{t-1}, \hat{\textbf{h}}_t), \\
    \textbf{P}_t &= Sample(\textbf{P}_{para, t}).
\end{flalign}

\emph{Upper Body Motion.} For upper body motion estimation, an ideal method is to build a body model and estimate the parameters \cite{XNect_SIGGRAPH2020}. To avoid making the algorithm too complex (upper body always lies only at the bottom part of the image), we assign the upper body part as a billboard \cite{cao2016real} which is shaped by several shoulder landmarks manually defined. The initial depth of the billboard is set as the average depth of the landmarks in full training sequences and same for all. We translate the billboard model with the 50\% translation part $T$ in the predicted head motion $P$ in most cases as the results.

\subsection{Photorealistic Image Synthesis}
\label{section 3.4}
The last stage of our approach is to generate the photorealistic facial renderings from previous predictions, as illustrated in Figure \ref{fig:pipeline}. Our rendering network is inspired by the recent advances in synthesizing photorealistic and controllable facial videos \cite{isola2017image, thies2019deferred, kim2018deep, elgharib2020egocentric}. We use a conditional image-to-image translation network as our backbone along with adversarial training. The network takes a channel-wise concatenation of a conditional feature map and $N = 4$ candidate images of the target person to produce photorealistic renderings.\\

\begin{figure}[t]
  \centering
  \includegraphics[scale = 0.3]{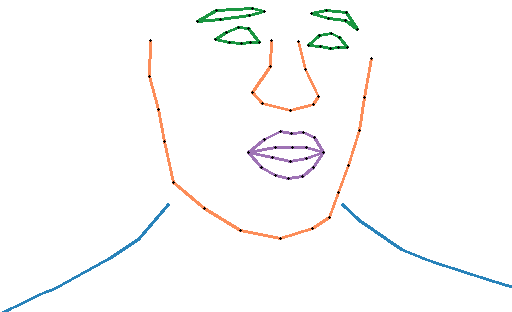}
  \caption{An example of our conditional feature map. Different colors are used to illustrate different semantic subsets. For example, green denotes the eyes and brows and blue denotes the upper body line. In practical experiments, the conditional feature map is monochrome, as shown in Figure \ref{fig:pipeline}.}
  \label{fig:conditional_map}
\end{figure}

\emph{Conditional Feature Maps.} To provide the clues of the face and the upper body, we draw a conditional feature map for each frame from the above predictions. An example of the conditional map is shown in Figure \ref{fig:conditional_map}. The feature map consists of a facial part and an upper body part. Drawing semantic regions with colors, or even further one region, one channel brings richer information and more drawing time. We didn’t find obvious improvements on these two alternatives. Note that both the sparse facial landmarks and the upper body billboard we predict locate in object coordinates. Therefore, we need to project these 3D positions to the 2D image plane via pre-computed camera intrinsic parameters $K$. The camera model we use is a pinhole camera model and $K = [f \ 0 \ c_x; 0 \ f \ c_y; 0 \ 0 \ 1]$, where $f$ is the focal length and $(c_x, c_y)$ is the principal point. The consecutive 2D projected components are line-connected in a pre-defined semantic sequence, resulting in the conditional feature map of size $1 \times 512 \times 512$.\\

\emph{Candidate Image set.} Besides the conditional feature map, we additionally input a candidate image set of the target person to provide detailed scene and texture clues. We found that adding such a candidate set helps the network generate consistent backgrounds considering the changing camera motions in the training set and relieves the pressure of the network to synthesize subtle details, like teeth and pores. These images are automatically selected. For the first two, we choose the 100th minimum/maximum mouth area. For the rest, we sample x- and y-axis rotation by uniform intervals and choose the nearest samples from intervals. Therefore, the size of the final concatenated input image becomes $13 (1 + 3 \times 4) \times 512 \times 512$. 

The network is a 8-layer UNet-like \cite{ronneberger2015u, esser2018variational, han2019finet} convolutional neural network with skip connections in each resolution layer. The resolution of each layer are (256$^2$, 128$^2$, 64$^2$, 32$^2$, 16$^2$, 8$^2$, 4$^2$, 2$^2$) and the corresponding numbers of feature channels are (64, 128, 256, 512, 512, 512, 512, 512). Each encoder layer consists of one convolution (stride 2) and one residual block. The symmetric decoder layer is almost the same, except the first convolution is replaced by a nearest up-sample operation with a scale factor of 2. Examples of our photorealistic renderings are shown in Figure \ref{fig:results_gallery}.

\section{Implementation Details}
In this section, we describe the aspects relevant to the implementation of our approach: dataset acquisition and pre-processing (Section \ref{section 4.1}), loss functions (Section \ref{section 4.2}), training setup (Section \ref{section 4.3}), and real-time animation settings (Section \ref{section 4.4}). 

\subsection{Dataset Acquisition and Pre-processing}
\label{section 4.1}

We apply our approach to 8 different target sequences of 7 different subjects for training and testing. These sequences span a range of 3-5 minutes. All videos are extracted at 60 frames per second (FPS) and the synchronized audio waves are sampled at 16$K$ Hz frequency. We first crop the video to keep the face at the center and then resize to $512 \times 512$. All input images and output images share the same resolution. We split videos as 80\% / 20\% for training and validation. Please refer to Appendix \ref{appendix} for more details.

We detect the 73 pre-defined facial landmarks for all videos using an off-the-shelf tool. To provide the groundtruth of the 3D mouth shape and head pose, we employ an optimization-based 3D face tracking algorithm similar to \cite{shi2014automatic, thies2016face2face}. For camera calibration, we use a binary search to compute the focal length $f$ as demonstrated in \cite{cao20133d}. The principle point $(c_x, c_y)$ is set as the center of the image. Note that we do camera calibration and 3D face tracking on the original image and compute a transformation matrix according to the crop and resize parameters. The upper body motion feature points are manually selected once for first frame of each sequence and tracked for the rest frames using LK optical flow \cite{bouguet2001pyramidal}, and the OpenCV implementation \cite{opencv_library}. For more details about monocular 3D face tracking, we recommend readers to refer to the summary paper \cite{zollhofer2018state}.

To train the APC speech representation extractor, we use the Mandarin Chinese part of the Common Voice dataset \cite{ardila2020common} that provides unlabelled wild utterances. Specifically, The subset contains 889 different speakers with various accents. In total, there are about 26 hours of unlabelled utterances. We use 80-dimensional log Mel spectrograms as surface features. The log mel spectrograms are computed with 1/60 second frame length, 1/120 second frame-shift, and 512-point Short-Time Fourier Transform (STFT). Although our APC model is trained in Mandarin, we find that our system still works well in other languages because the model learns the high-level and semantic information. Also the manifold projection improves the generalization ability.

\subsection{Loss Functions}
\label{section 4.2}
\subsubsection{Deep Speech Representation Extraction} 
The training of the APC model is fully self-supervised via predicting the surface features $n$ frames ahead. Given a sequence of log mel spectrograms $(x_1, x_2, ..., x_T)$, the APC model processes each element $x_t$ at time step $t$ and outputs a prediction $y_t$, generating a predicted sequence $(y_1, y_2, ..., y_T)$. We optimize the model by minimizing the L1 loss between the input sequence and the predicted as follows:
\begin{equation}
    \sum_{i=1} ^{T-n} |x_{i+n} - y_i|,
\end{equation}
where $n = 3$ following the setting in \cite{APC}.

\subsubsection{Audio to Mouth-related Motion} 
To learn the mapping from audio to mouth-related motion, we minimize the $L_2$ distance between the groundtruth mouth displacements and the predicted displacements. Specifically, the loss can be written as:
\begin{equation}
    \sum_{t=1}^T \sum_{i=1}^N || \Delta \textbf{v}_{m,t} - \Delta \hat{\textbf{v}}_{m,t} ||_2^2,
\end{equation}
where $T = 240$ represents the number of consecutive frames sent to the model at each iteration. $N = 25$ is the number of the pre-defined sparse mouth-related 3D points in our experiments.

\subsubsection{Probabilistic Head Motion Synthesis} 
Apart from learning the mapping from audio to mouth-related motion, we also aim to estimate the target's head pose during training. The upper body motion can be deduced from the head pose as mentioned in Section \ref{section 3.3}. Specifically, we employ an autoregressive probabilistic model to model the head pose distribution. We train the model by minimizing the negative log-likelihood of the pose distribution. Given a sequence of history head pose $(x_{t-F}, ..., x_t)$ and speech representation $\textbf{h}_t$, the probabilistic loss is:
\begin{equation}
    -\ln (\mathcal{N}(\textbf{x}_t, \textbf{h}_t|\hat{\mu}_n, \hat{\sigma}_n)),
\end{equation}
where $\textbf{x}_t, \textbf{h}_t$ is the input head pose and speech representation at time $t$. This loss term forces the model to output the mean values $\hat{\mu}_n$ and standard deviations $\hat{\sigma}_n$ of the Gaussian distribution. To increase numerical stability, we output the negative log sigma instead of the sigma directly. Each element $\textbf{x}_t \in \mathbb{R}^{12}$ in pose sequence are composed of the current pose $\textbf{p}_t \in \mathbb{R}^{6}$ and a linear velocity term $\Delta \textbf{p}_t \in \mathbb{R}^{6}$. Although we only use the first six dimensions of rotation and translation after sampling in the distribution, we find that adding such a velocity term could implicitly force the model to focus on the motion speed, leading to smoother results.

\subsubsection{Photorealistic Image Synthesis}
Finally, we train a neural renderer to synthesis photorealistic talking-head images. The training procedure follows the adversarial training mechanism \cite{goodfellow2014generative}. We adopt the multi-scale PatchGAN architecture \cite{isola2017image, wang2018high} as the backbone of discriminator $\textbf{D}$. The image-to-image translation network $\textbf{G}$ is trained to generate "realistic" images to fool the discriminator $\textbf{D}$, while the discriminator $\textbf{D}$ is trained to tell the generated images from groundtruth images. Specifically, we employ LSGAN loss \cite{mao2017least} as the adversarial loss to optimize the discriminator $\textbf{D}$:
\begin{flalign}
    \mathcal{L}_{GAN}(\textbf{D}) =  (\hat{r} - 1)^2 + r^2, 
\end{flalign}

where $\hat{r}, r$ is the discriminator classification output when input the groundtruth image $\hat{y}$ and the generated rendering $y$, respectively. We additionally use a color loss, a perceptual loss \cite{johnson2016perceptual} and a feature matching loss \cite{wang2018high}:

\begin{flalign}
     \mathcal{L}_{G}  = \mathcal{L}_{GAN}(\textbf{G}) + \lambda_C \mathcal{L}_{C} + \lambda_P \mathcal{L}_{P} + \lambda_{FM} \mathcal{L}_{FM},
\end{flalign}
where $\mathcal{L}_{GAN}(\textbf{G}) = (r - 1)^2 $ is the adversarial loss, $\mathcal{L}_{C}$ is the color loss, $\mathcal{L}_{P}$ is the perceptual loss and $\mathcal{L}_{FM}$ is the feature matching loss. The weights $\lambda_C, \lambda_P, \lambda_{FM}$ of each loss are set to $(100, 10, 1)$ in all our experiments empirically. The color loss is a $L_1$ per-pixel loss to minimize the difference from generated images $y$ and groundtruth images $\hat{y}: \mathcal{L}_{C} = || y - \hat{y} ||_1.$ We tried higher weights (x10) on mouth, yep mouth-related errors drop, but full-image errors rise. Considering the full-image generation task, we choose equal weight.

\begin{figure*}[h]
    \centering
  \includegraphics[width=\linewidth, scale=0.1]{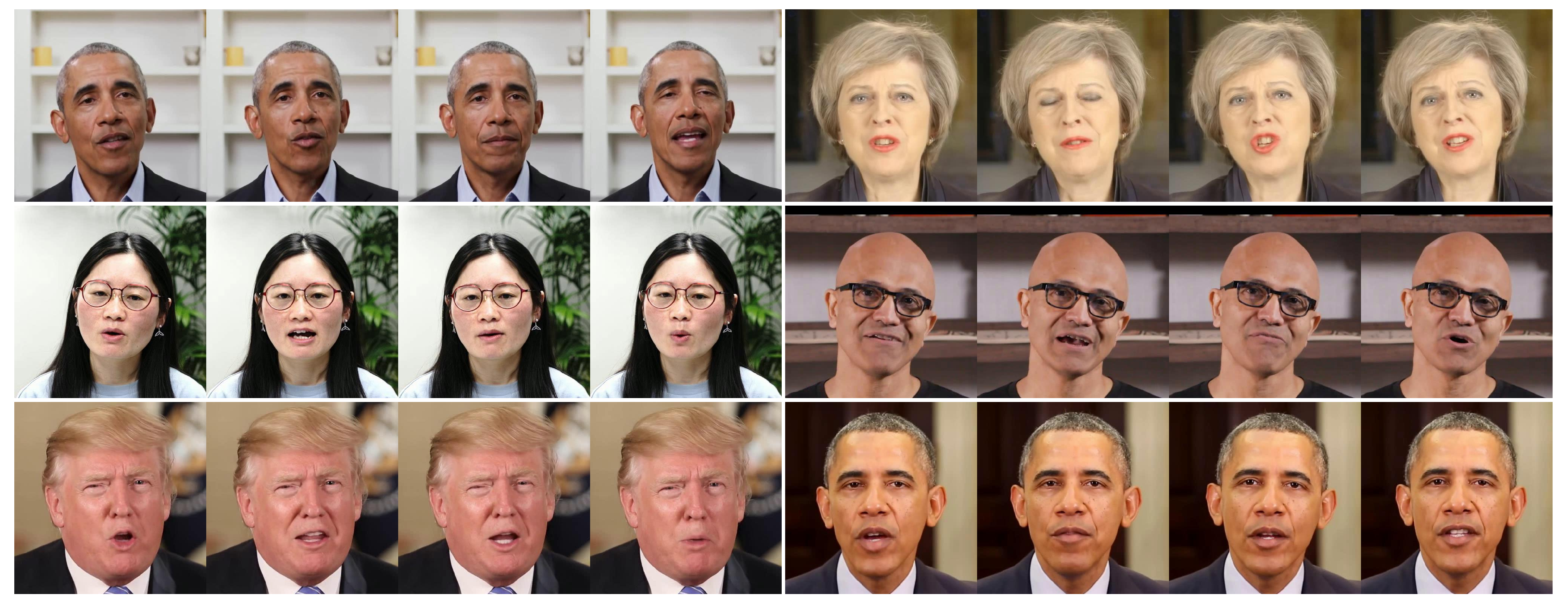}
  \caption{A gallery of our audio-driven talking-head animation results. Given an arbitrary audio stream, our method first generates personalized facial dynamics, head poses, and upper body motions and then synthesizes photorealistic renderings from these predictions. Please refer to the supplementary video for full sequences.
  Video (at upper left corner) \emph{Obama} \copyright \emph{Barack Obama Foundation} (public domain). Video \emph{May} \copyright \emph{UK government} (Open Government Licence). Video \emph{Nadella} \copyright \emph{IEEE Computer Society} (public domain). Video \emph{Trump} \copyright \emph{White House} (public domain). Video (at bottom right corner) \emph{Obama} \copyright \emph{White House} (public domain).}
  \label{fig:results_gallery}
\end{figure*}

For perceptual loss, we adopt a VGG19 network \cite{simonyan2014very} to extract the perceptual features from $\hat{y}$ and $y$ and minimize their $L_1$ distance:
\begin{equation}
    \mathcal{L}_{P} =  \sum_{i \in \mathcal{S}} || \phi^{(i)}(y) - \phi^{(i)}(\hat{y}) ||_1,
\end{equation}
where $\mathcal{S} = \{1, 6, 11, 20, 29\}$ denotes the layers we use and $\phi^{(i)}$ denotes the $i$-th layer. Finally, to improve the training speed and stability, we adopt a feature matching loss:
\begin{equation}
    \mathcal{L}_{FM} = \sum_{i=1}^L || r - \hat{r} ||_1
\end{equation}
where $L$ is the number of spatial layers in discriminator $\textbf{D}$. The $L_1$-based feature matching loss is designed to match the statistics of features extracted by the discriminator from $\hat{y}$ and $y$.

\subsection{Training Setup and Parameters}
\label{section 4.3}
All our models are trained on PyTorch (Python) \cite{NEURIPS2019_9015} using Adam optimizer with hyper-parameters $(\beta_1, \beta_2) = (0.9, 0.999)$. The learning rate is set to $10^{-4}$ and linearly decay to $10^{-5}$ in all experiments. The APC model contains 4.064$M$ parameters, the mouth-related position predictor contains 3.064$M$ parameters, the head pose estimator contains 4.267$M$ parameters and the renderer contains 76.204$M$ parameters. 

We train the first three models on a Nvidia 1080Ti GPU and takes about (11, 0.5, 5) hours in total (200, 200, 200) epochs with a batch size of 32, respectively. The photorealistic images renderer is trained on 4 Nvidia 1080Ti GPUs for an average of 22 hours in 60 epochs and a batch size of 8. During testing, we select all models with minimized validation loss.

\subsection{Real-Time Animation}
\label{section 4.4}
We implement and test our real-time animation system using C++ on a desktop PC with an Intel Core i7-9700K CPU (32 GB RAM) and an NVIDIA GeForce RTX 2080 (8 GB RAM). The deep speech representation extraction module takes around 2.9 ms for inference (1.4 ms for APC model forward pass and 1.5 ms for manifold projection). Learning facial dynamics from audio representations via 3-layer LSTM and MLP networks takes around 2.5 ms. Besides, the Audio2Mouth module leads to around 300 ms latency for obtaining 18 frames of future audio information. We then use Tensorrt to accelerate the last two models. Specifically, the head poses estimation model takes 4.4 ms and the photorealistic renderer takes 20.1 ms after acceleration. Time of memory copying between CPU and GPU has already been included here. Therefore, the entire system takes about 27.4 ms for inference at over 30 FPS with 300 ms latency.\\

\emph{Discussion.} Here we discuss the run time with related work Zhou et al. \shortcite{makelttalk} and Thies et al. \shortcite{NVP}. We emphasize that our system is the first \emph{realization} of an end-to-end live system for photorealistic audio-driven talking-head animation and step further to practical applications, considering that these papers haven't shown an actual live demo. Besides, Zhou et al. \shortcite{makelttalk} is not designed for live streaming generation. The self-attention network \cite{transformer} in speaker-aware animation works as a post weighted combination of previous landmarks predictions and is inappropriate for live applications like video conferencing, requiring a low latency. Thies et al. \shortcite{NVP} will meet more difficulties. Their method is restricted to the target video length, and therefore a sufficiently long target video without obstruction is required, which is hard to acquire. To generate longer frames, an additional heuristic schedule to select proper candidate frames is needed. Also, pose-audio inconsistency appears for the lacking control on head motion (Section \ref{section 2}). These factors result in obstacles in the live implementation.

\begin{figure}[h]
    \centering
  \includegraphics[width=\linewidth, scale=1]{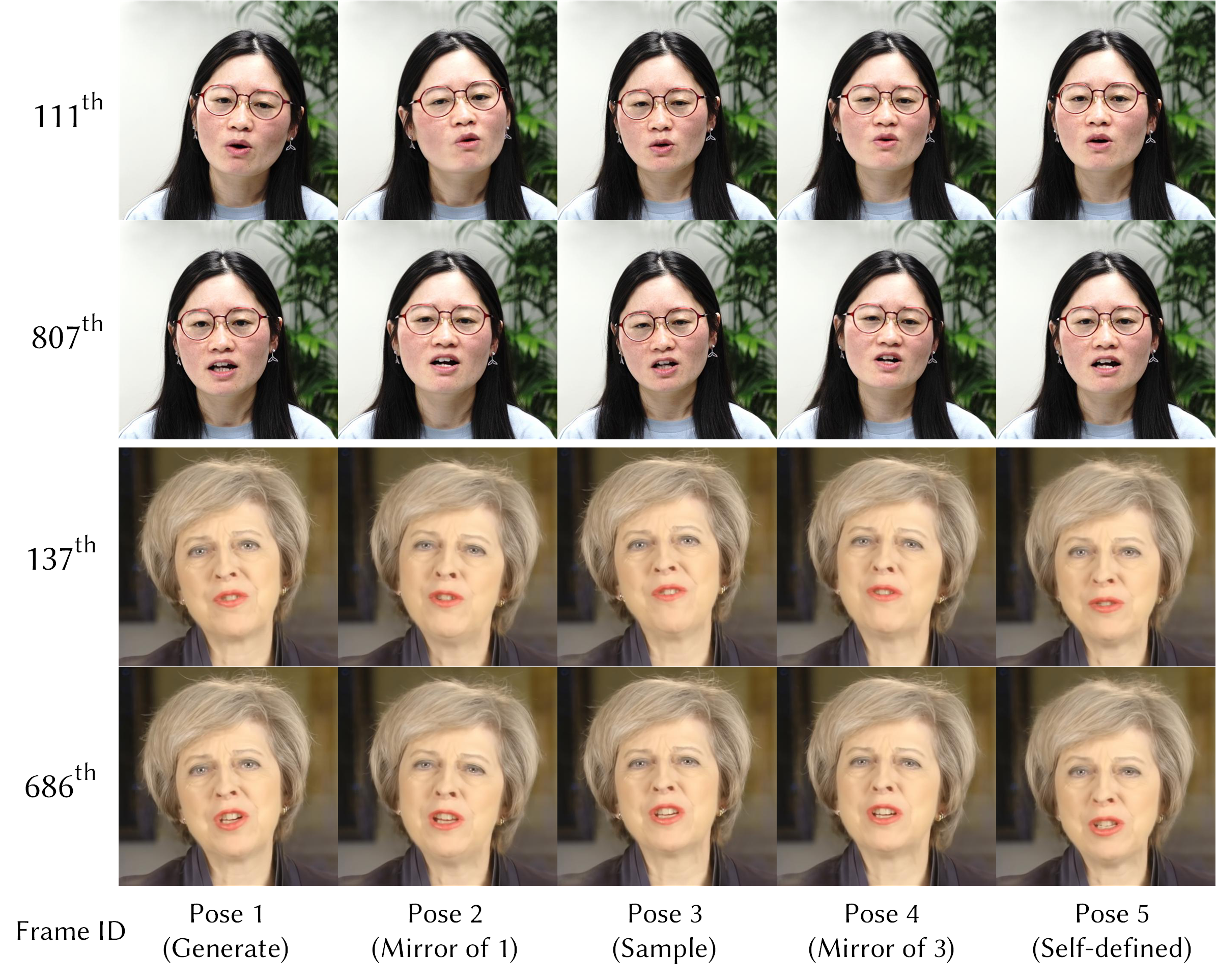}
  \caption{Our method allows generating pose controllable results. We demonstrate results with different poses here. The number at the left of the image denotes the frame index of the video. Pose 1 is generated by our model, and Pose 3 is sampled from the training set. To further evaluate the head poses controllability, we mirror the head poses 1 and 3 to generate head pose 2 and 4. Pose 5 is self-defined. Please refer to the supplementary video for full sequences.
  Video \emph{May} \copyright \emph{UK government} (Open Government Licence).}
  \label{fig:results_headpose}
\end{figure}

\begin{figure}[h]
  \centering
  \includegraphics[width=\linewidth]{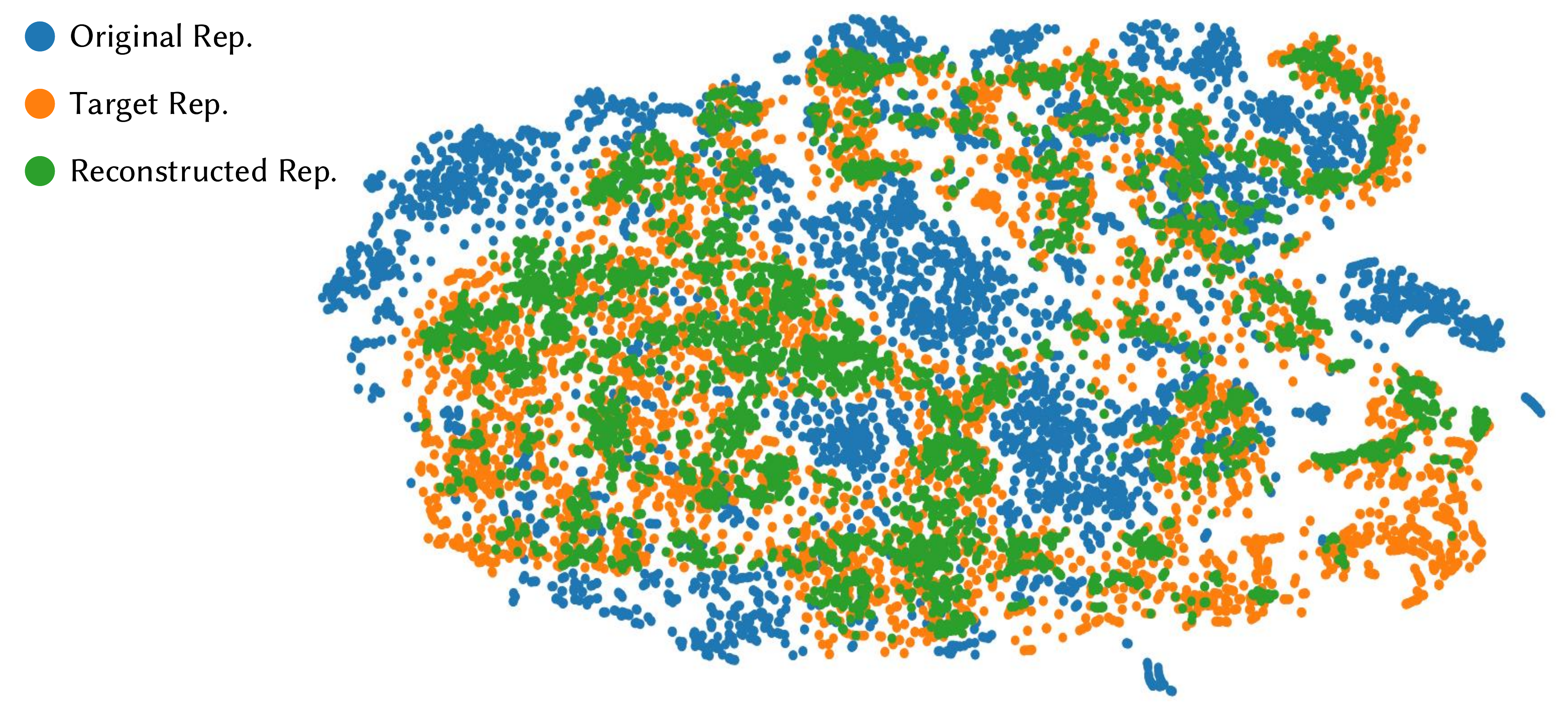}
  \caption{t-SNE visualization of Manifold Projection. As shown in the legend, original, target and reconstructed representations are marked using different colors.}
  \label{fig:manifoldprojection_vis}
\end{figure}

\begin{figure}[h]
  \centering
  \includegraphics[width=\linewidth]{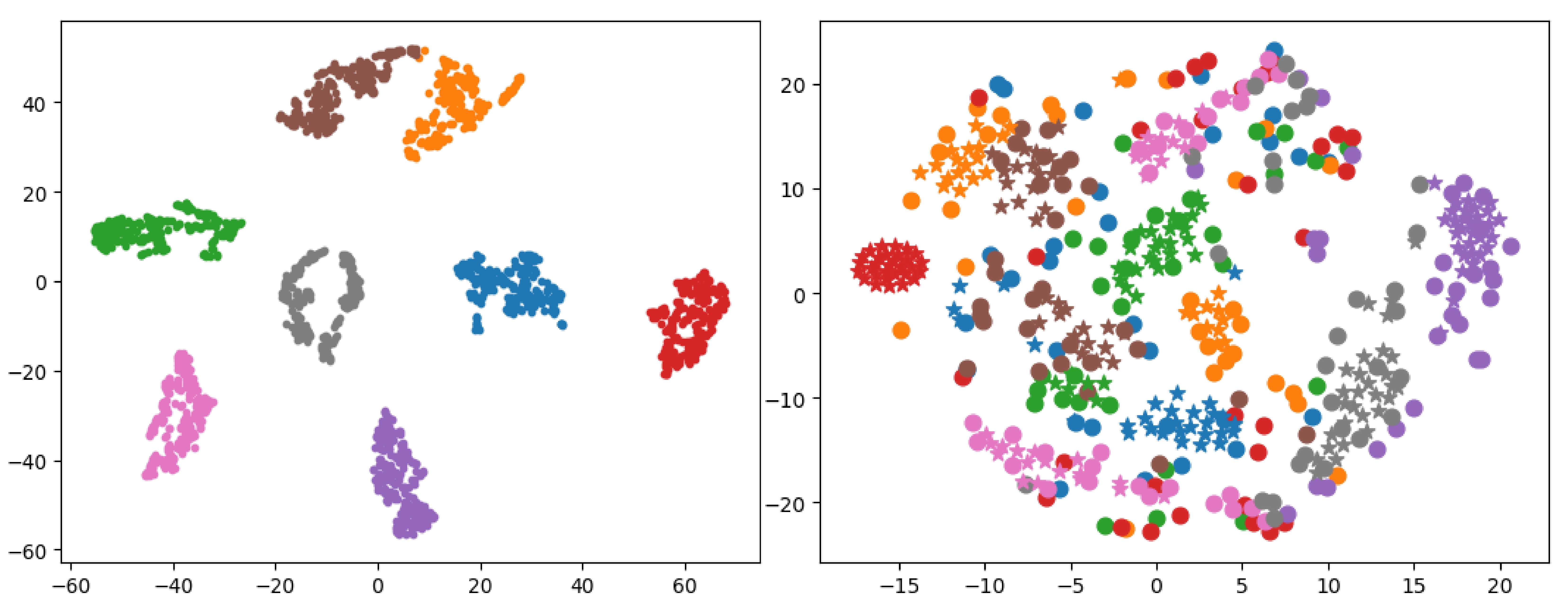}
  \caption{t-SNE visualization of head pose generation. Different targets are marked using different colors. Left: visualization of generated poses. Right: visualization of generated poses and head poses from training corpus. Marker \emph{solid dot} denotes training corpus poses and marker \emph{star} denotes generated poses.}
  \label{fig:pose_vis}
\end{figure}

\section{Results}
Our live speech portraits method generates personalized and photorealistic talking-head animation from audio input in real-time. We recommend readers watch the supplementary video.

\begin{figure}[h]
  \centering
  \includegraphics[width=\linewidth]{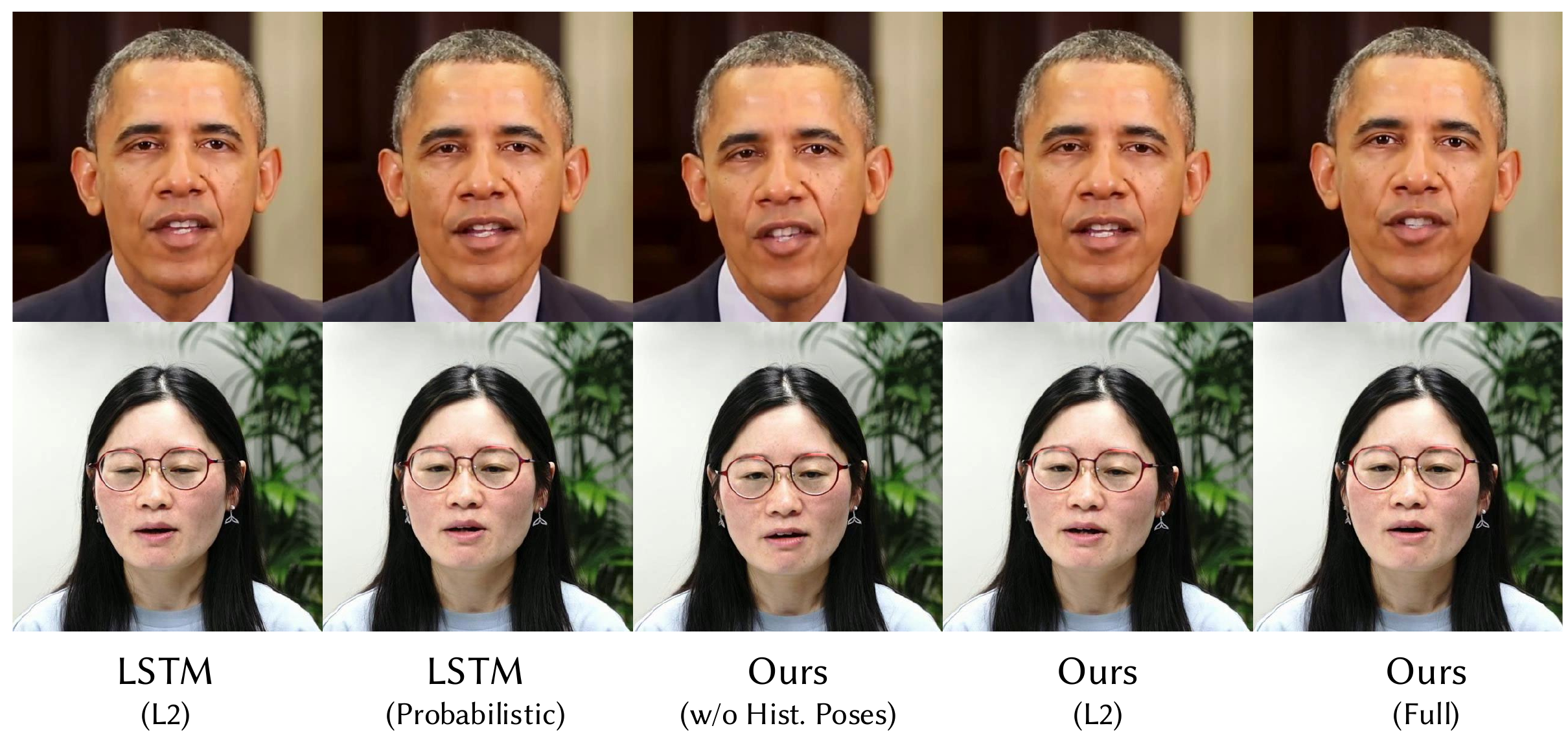}
  \caption{Qualitatively evaluation of head pose estimation. We compare our model design with several alternative variants. Please refer to the supplementary video for full sequences.
  Video \emph{Obama} \copyright \emph{White House} (public domain).}
  \label{fig:ab_headpose}
\end{figure}

In the following, we present the results of our approach, evaluate the design of our method both qualitatively and quantitatively, compare to the state-of-the-art techniques, and show results of a user study. We further demonstrate the potentials on several applications, e.g., dubbing, video conferencing and virtual avatars.

\subsection{Qualitative Evaluation}
\label{section qualitative}
Figure \ref{fig:results_gallery} shows a gallery of our results. Our approach allows animating target portraits driven by audio sequences while preserving the personal talking styles. Our method produces facial dynamics, natural head poses, and upper body motions and synthesizes temporally coherent renderings in high-fidelity, e.g., clear teeth textures. It also successfully works on subjects with long hair, glasses, and earrings. Figure \ref{fig:teaser} and the supplementary video show examples of our real-time animation system and results of different people driven by the same utterance. Results prove that we preserve their own talking styles in the training video.

Figure \ref{fig:results_headpose} demonstrates the pose controllability of our method. We synthesize results with different head poses, either generated by our system or sampled from the training set (Columns 1 and 3). We further evaluate our model with mirrors of the previous two poses and a self-defined pose (Columns 2, 4, and 5). Even though these challenging poses partly lay outside the training corpus and therefore may lead to artifacts, our system still generates correct lip motions and temporally smooth renderings.

We analyze the effectiveness of manifold projection in Figure \ref{fig:manifoldprojection_vis}. The operation is designed to project the original speech representations to the target speech space by minimizing the reconstruction loss. As shown in the figure, the reconstructed features (green) are much closer to the target speech space than the original wild features (blue). More quantitative comparison results of the manifold projection can be seen in our supplementary video. We found that using manifold projection generates more accurate lip synchronization than not using it, especially when encountering the audio of different genders or foreign languages.

Figure \ref{fig:pose_vis} depicts a qualitative evaluation of the generated head poses using t-SNE visualization. We select 8 different targets and denote them with different colors. In Figure \ref{fig:pose_vis} (left), we visualize the generated head poses of these targets from the same input audio. It can be seen that our predictions for one person lie in a nearby region and are far away from different targets. We further analyze the connections between generated poses and training poses in Figure \ref{fig:pose_vis} (right). \emph{Solid dot} denotes training corpus poses and \emph{star} denote our predictions. We downsample the training poses using K-means to 30 to reduce data size. The visualization demonstrates that our predictions locate closely to training corpus poses.

\begin{table}[]
    \caption{Quantitative evaluation of input time delays. Validation losses are computed as Euclidean distance of landmarks.}
    \label{table_latency}
    \begin{tabular}{c|c|c|c|c}
    \hline
    \hline
    Time Delay   & 0ms & 50ms & 100ms & 150ms \\ \hline 
    Val Loss (mm)    & 5.309  &   5.280  &  5.117 & 5.335  \\ \hline \hline
    Time Delay   & 200ms &  300ms  &  400ms  &  500ms      \\ \hline
    Val Loss (mm) &  5.160   &   \textbf{4.916}  &  5.248 & 5.539    \\ \hline
    \end{tabular}
\end{table}

We also qualitatively evaluate the design of the head pose estimation module from audio. Results are shown in Figure \ref{fig:ab_headpose}. We recommend readers to watch the supplementary video for better visual comparisons. To confirm our proposed two prior assumptions, here we perform an ablative study of the input, architecture, and loss design by training and testing four alternative variants: "\emph{LSTM (L2)}" (LSTM network trained using L2 loss), "\emph{LSTM (Probabilistic)}" (LSTM network trained using probabilistic loss), "\emph{Ours (w/o Hist. Poses)}" (Our architecture without history poses input), "\emph{Ours (L2)}" (Our architecture trained using L2 loss). A quantitative evaluation of head pose estimation can be found in Section \ref{section 5.2}.

The variants "\emph{LSTM (L2)}" and "\emph{LSTM (Probabilistic)}" generate more temporal jitters than other variants, which indicates that directly using LSTM architecture may not be a good choice in our task. A reasonable explanation is that RNN is easy to overfit on a small dataset, which is an around 3-minute length video in our case. On the other hand, LSTMs have a theoretically infinite receptive field of the history information depending on the forget and memory mechanism. Our architecture has a fixed receptive field which intuitively becomes more robust to long history information and will not easily overfit.

The variant "\emph{Ours (w/o Hist. Poses)}" also performs worse than our full model. It tends to generate unchanged poses in the \emph{Obama} video and struggled poses in the woman video (see the supplementary video). Without modeling the history poses, the model is trained to learn a mapping from audio to the head poses solely. Therefore, the network generates pose with the most probability of the current audio clip, which may be far away from adjacent ones, leading to time-incoherent results.

The variant "\emph{Ours (L2)}" generates best results among all variants but there is still room for improvement in the term of time consistency and motion variation. The only difference is that this variant replaces the probabilistic loss with the L2 distance loss. It makes the problem a regression problem, which means the model needs to find the deterministic position given history poses and audio information. Therefore, the model struggles to find the best balance point between audio and history poses, resulting in time-inconsistency and minor motion variations. A probabilistic model is suitable to handle these ambiguities and finally achieves the best results both in time consistency and motion richness.

\begin{table}[]
    \caption{Quantitative evaluation of head pose prediction. We compare our approach with alternative methods in Section \ref{section qualitative} and Zhou et al. \shortcite{makelttalk}. $\downarrow$ denotes lower is better.}
    \label{table_headpose}
    \begin{tabular}{c|c|c|c}
    Methods                & D-L $\downarrow$ & D-V $\downarrow$ & D-Rot/Pos $\downarrow$ \\ \hline \hline
    LSTM (L2)               &  4.9\%  &   1.1\%  &   6.9/12.2\%        \\ \hline
    LSTM (Probabilistic)   &   4.9\%  &  1.0\%   &     6.7/11.6\%      \\ \hline
    Ours (w/o Hist. Poses) &  3.9\%   &   0.9\%  &       4.2/8.9\%    \\ \hline
    Ours (L2)              &  4.5\%   &  1.1\%   &         3.7/9.2\%  \\ \hline \hline
    \cite{makelttalk}      &  4.6\%   &  0.9\%   &        6.1/10.1\%   \\ \hline
    Ours (Full)             &  \textbf{3.6\%}   &  \textbf{0.8\%}   &      \textbf{3.6}/\textbf{8.9\%}   
    \end{tabular}
\end{table}

\subsection{Quantitative Evaluation}
\label{section 5.2}
Here, we perform a thorough quantitative evaluation of the proposed method. Table \ref{table_latency} evaluates the importance of time latency of input audio. Choosing a proper latency is important for a real-time system because it greatly affects the user experience. In this experiment, we train the audio-to-mouth network with different time delays on a 30-minutes subset of the \emph{Obama Weekly Address} dataset \cite{SIG17Obama}. The subset was split into a training dataset (80\%) and a validation dataset (20\%). Validation losses are computed as Euclidean distance between predicted mouth-related landmarks positions and the tracked groundtruth. As can be seen, the time delay of 300 ms (18 frames) gives minimized validation loss. Too short or too long delays both lead to performance reduction. When the network fails to have access to a certain range of future information (smaller than 300 ms), it cannot model the coarticulation. Too long delays (longer than 300 ms) always cover several phonemes and introduce redundant information. In most real-time scenarios, e.g., video conferencing, 300 ms is a tolerable latency. Therefore, we use this latency in all experiments.

Now we perform quantitative evaluation of our head pose estimation module. We compare our method with using four alternative variants pre-mentioned in Section \ref{section qualitative} as well as Zhou et al. \shortcite{makelttalk}. We created a test set composed of 6 different videos (average lasts for 45 seconds each clip) and the speeches and videos are unseen during training. Specifically, We evaluate these methods on the testing set by computing the metrics \textbf{D-L}, \textbf{D-V} and \textbf{D-Rot/Pos} defined in \cite{makelttalk}. \textbf{D-L} and \textbf{D-V} denote the normalized Euclidean position and velocity difference between predictions and the groundtruth. \textbf{D-Rot/Pos} denotes the rotation angle differences and normalized translation distances. Lower is better for all these metrics.

\begin{figure}[h]
  \centering
  \includegraphics[width=\linewidth]{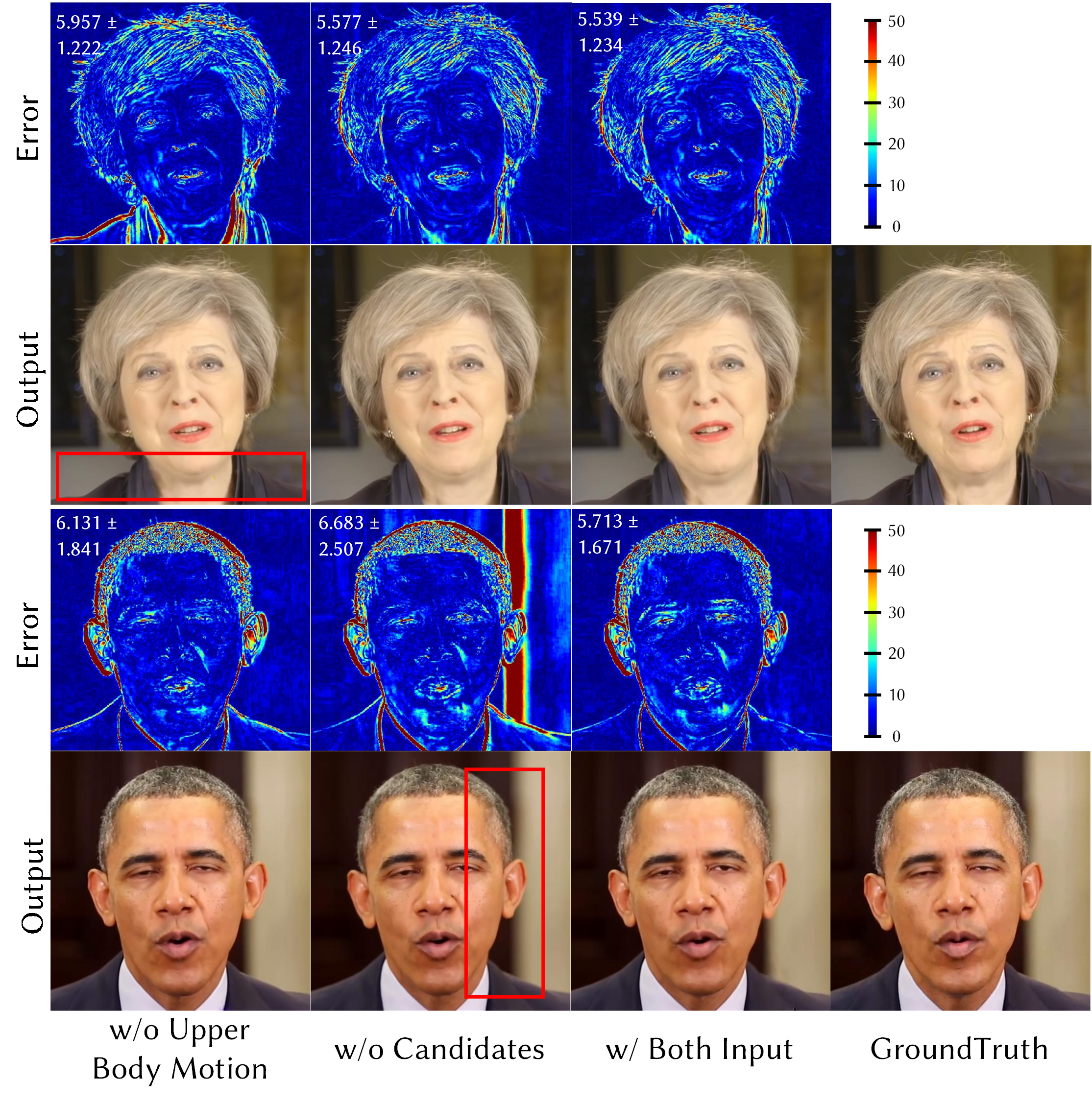}
  \caption{Quantitative evaluation of the conditional input of the renderer.
  Video \emph{May} \copyright \emph{UK government} (Open Government Licence). Video \emph{Obama} \copyright \emph{White House} (public domain).}
  \label{fig:ab_input}
\end{figure}

Table \ref{table_headpose} reports the evaluation results. Our model obviously outperforms other alternative variants, tallying with the subjective evaluation in Section \ref{section qualitative}. In particular, we observe that the LSTM-based variants produce higher error than other variants ((see "\emph{LSTM (L2)}" and "\emph{LSTM (Probabilistic)}")). This results confirms that LSTM is prone to overfit in the training set. Replacing the probabilistic modeling with L2-regression generates slightly worse results (see "\emph{Ours (L2)"}). Our full method generates similar head motion with respect to the groundtruth, confirming that we learn the distribution of the target person. We also compare with Zhou et al. \shortcite{makelttalk} which learn speaker-aware head motions. Results demonstrate that our method outperforms their method.

\begin{figure}[h]
  \centering
  \includegraphics[width=\linewidth]{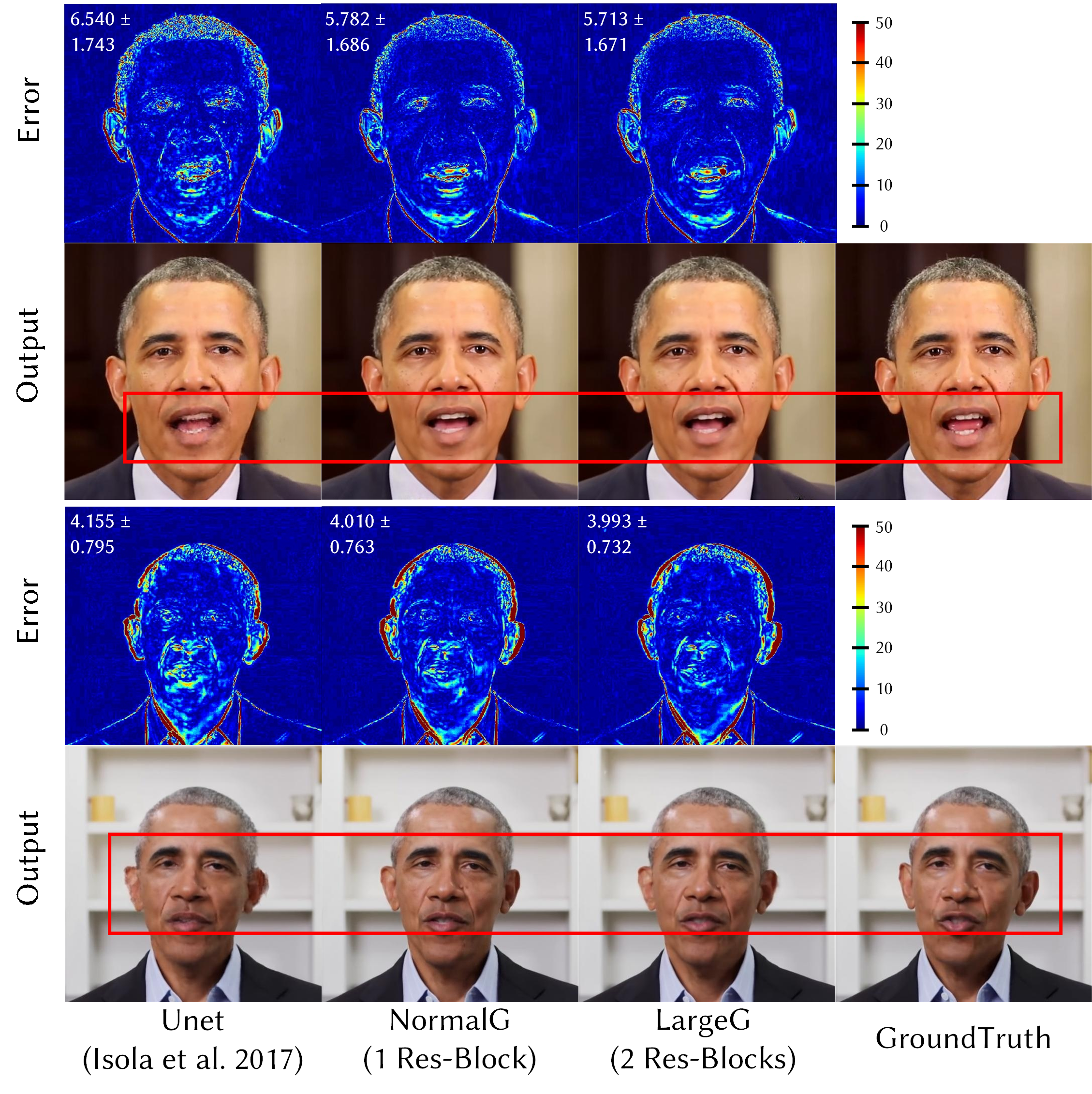}
  \caption{Quantitative evaluation of the architecture design of the renderer.
  Video (Upper) \emph{Obama} \copyright \emph{White House} (public domain). Video (Lower) \emph{Obama} \copyright \emph{Barack Obama Foundation} (public domain).}
  \label{fig:ab_architecture}
\end{figure}

\begin{table*}[]
    \caption{Quantitative evaluation of photorealistic renderer design. We evaluate the results of input condition (top), architecture (middle), and training dataset size (bottom). $\downarrow$ denotes lower is better and  $\uparrow$ denotes higher is better.}
    \label{table_renderer}
    \begin{tabular}{c|c|c|c|c|c}
    \hline
    \multicolumn{2}{c|}{Target/Term/Method} & L1 $\downarrow$ & PSNR $\uparrow$ & SSIM $\uparrow$ & LPIPS $\downarrow$ $\times$ 10 \\ \hline
    \multirow{3}{*}{Obama1} & w/o U.B.M. & 6.131 $\pm$ 1.841 & 25.364 $\pm$ 2.259 & 0.850 $\pm$ 0.043 & 0.704 $\pm$ 0.407 \\ \cline{2-6} 
    & w/o Cand. & 6.683 $\pm$ 2.507& 24.964 $\pm$ 2.483 & 0.845 $\pm$ 0.048 & 0.746 $\pm$ 0.422 \\ \cline{2-6} 
    & w/ Both & \textbf{5.713} $\pm$ 1.671 & \textbf{26.006} $\pm$ 2.275 & \textbf{0.862} $\pm$ 0.038 & \textbf{0.698} $\pm$ 0.348  \\ \hline
    \multirow{3}{*}{Obama2} & w/o U.B.M. & 7.634 $\pm$ 5.105 & 23.143 $\pm$ 2.758 & 0.843 $\pm$ 0.102 & 1.136 $\pm$ 1.279 \\ \cline{2-6} 
    & w/o Cand. & 4.006 $\pm$ 0.754 & \textbf{27.927} $\pm$ 2.228 & 0.917 $\pm$ 0.020 & \textbf{0.366} $\pm$ 0.111 \\ \cline{2-6} 
    & w/ Both & \textbf{3.993} $\pm$ 0.732 & 27.576 $\pm$ 2.174 & \textbf{0.926} $\pm$ 0.020  & 0.370 $\pm$ 0.103 \\ \hline
    \multirow{3}{*}{May} & w/o U.B.M. & 5.957 $\pm$ 1.322 & 27.171 $\pm$ 1.820 & 0.807 $\pm$ 0.037 & 0.790 $\pm$ 0.260 \\ \cline{2-6} 
    & w/o Cand. & 5.577 $\pm$ 1.246 & 27.979 $\pm$ 1.683  & 0.823 $\pm$ 0.033 & 0.747 $\pm$ 0.280 \\ \cline{2-6} 
    & w/ Both & \textbf{5.539} $\pm$ 1.234 & \textbf{28.044} $\pm$ 1.782 & \textbf{0.823} $\pm$ 0.035 & \textbf{0.746} $\pm$ 0.209 \\ \hline
    \end{tabular}

    \begin{tabular}{c|c|c|c|c|c}
    \multicolumn{2}{c|}{Target/Term/Method} & L1 $\downarrow$ & PSNR $\uparrow$ & SSIM $\uparrow$ & LPIPS $\downarrow$ $\times$ 10 \\ \hline
    \multirow{3}{*}{Obama1} & Unet & 6.540 $\pm$ 1.743 & 25.087 $\pm$ 2.046 & 0.832 $\pm$ 0.035 & 0.875 $\pm$ 0.395 \\ \cline{2-6} 
    & Normal & 5.782 $\pm$ 1.687 & 25.920 $\pm$ 2.229 & 0.861 $\pm$ 0.038 & 0.727 $\pm$ 0.346 \\ \cline{2-6} 
    & Large & \textbf{5.713} $\pm$ 1.671 & \textbf{26.006} $\pm$ 2.275 & \textbf{0.862} $\pm$ 0.038 & \textbf{0.698} $\pm$ 0.348  \\ \hline
    \multirow{3}{*}{Obama2} & Unet & 4.155 $\pm$ 0.794 & 27.414 $\pm$ 2.182 & 0.907 $\pm$ 0.022 & 0.479 $\pm$ 0.124 \\ \cline{2-6} 
    & Normal & 4.010 $\pm$ 0.763 & 27.406 $\pm$ 2.211 & 0.912 $\pm$ 0.021 & 0.376 $\pm$ 0.106 \\ \cline{2-6} 
    & Large & \textbf{3.993} $\pm$ 0.732 & \textbf{27.576} $\pm$ 2.174 & \textbf{0.926} $\pm$ 0.020  & \textbf{0.370} $\pm$ 0.103 \\ \hline
    \multirow{3}{*}{May} & Unet & 6.007 $\pm$ 1.709 & 27.528 $\pm$ 1.835 & 0.806 $\pm$ 0.037 & 1.018 $\pm$ 0.389 \\ \cline{2-6} 
    & Normal & 5.578 $\pm$ 1.694 & 27.924 $\pm$ 1.882  & 0.818 $\pm$ 0.049 & 0.828 $\pm$ 0.449 \\ \cline{2-6} 
    & Large & \textbf{5.539} $\pm$ 1.234 & \textbf{28.044} $\pm$ 1.782 & \textbf{0.823} $\pm$ 0.035 & \textbf{0.746} $\pm$ 0.209 \\ \hline
    \end{tabular}

    \begin{tabular}{c|c|c|c|c|c}
    \multicolumn{2}{c|}{Target/Term/Method} & L1 $\downarrow$ & PSNR $\uparrow$ & SSIM $\uparrow$ & LPIPS $\downarrow$ $\times$ 10 \\ \hline
    \multirow{3}{*}{Obama1} & 0.5 mins & 9.403 $\pm$ 3.068 & 22.481 $\pm$ 2.166 & 0.788 $\pm$ 0.042 & 1.421 $\pm$ 0.533 \\ \cline{2-6} 
    & 1 min & 8.259 $\pm$ 2.552 & 23.939 $\pm$ 2.427 & 0.812 $\pm$ 0.046 & 1.233 $\pm$ 0.523 \\ \cline{2-6} 
    & 3 mins & \textbf{5.713} $\pm$ 1.671 & \textbf{26.006} $\pm$ 2.275 & \textbf{0.862} $\pm$ 0.038 & \textbf{0.698} $\pm$ 0.348  \\ \hline
    \multirow{3}{*}{Obama2} & 0.5 mins & 6.941 $\pm$ 2.015 & 22.960 $\pm$ 2.195 & 0.858 $\pm$ 0.030 & 0.945 $\pm$ 0.382 \\ \cline{2-6} 
    & 1 min & 5.285 $\pm$ 1.303 & 24.873 $\pm$ 2.010 & 0.888 $\pm$ 0.024 & 0.585 $\pm$ 0.239 \\ \cline{2-6} 
    & 3 mins & \textbf{3.993} $\pm$ 0.732 & \textbf{27.576} $\pm$ 2.174 & \textbf{0.926} $\pm$ 0.020  & \textbf{0.370} $\pm$ 0.103 \\ \hline
    \multirow{3}{*}{May} & 0.5 mins & 9.539 $\pm$ 2.740 & 23.432 $\pm$ 2.087 & 0.739 $\pm$ 0.033 & 1.731 $\pm$ 0.582 \\ \cline{2-6} 
    & 1 min & 7.265 $\pm$ 2.366 & 25.875 $\pm$ 2.086  & 0.739 $\pm$ 0.050 & 1.175 $\pm$ 0.685 \\ \cline{2-6} 
    & 3 mins & \textbf{5.539} $\pm$ 1.234 & \textbf{28.044} $\pm$ 1.782 & \textbf{0.823} $\pm$ 0.035 & \textbf{0.746} $\pm$ 0.209 \\ \hline
    \end{tabular}
\end{table*}

\begin{figure}[h]
  \centering
  \includegraphics[width=\linewidth]{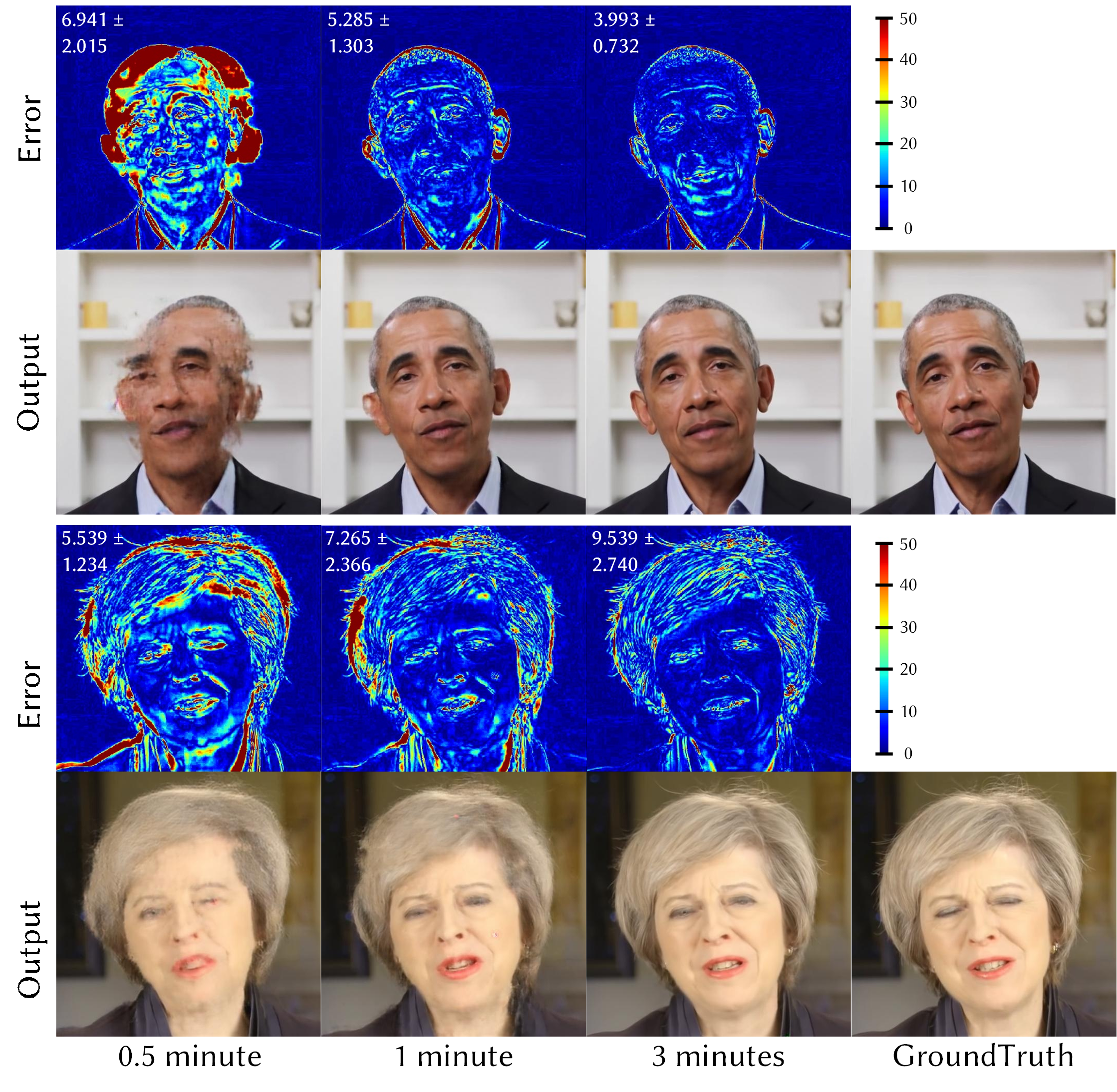}
  \caption{Quantitative evaluation of training dataset size.
  Video \emph{Obama} \copyright \emph{Barack Obama Foundation} (public domain). Video \emph{May} \copyright \emph{UK government} (Open Government Licence).}
  \label{fig:ab_datasize}
\end{figure}

Here, we quantitatively evaluate the photorealistic renderer. Results can be found in Figure \ref{fig:ab_input}-\ref{fig:ab_datasize}. We evaluate the model on three videos. The first three minutes of these videos are used as training data while the rest work as the testing set unless stated otherwise. Note that we test models with groundtruth head poses for numerical evaluation. We report the numerical errors and corresponding standard deviation of average L1-photometric loss (range of 0-255), Peak Signal to Noise Ratio (PSNR), Structural Similarity (SSIM), and deep perceptual distance (LPIPS) \cite{Zhang_2018_CVPR} in Table \ref{table_renderer}. The standard deviation indicates the temporal stability around the mean loss - higher variance means stronger incoherence. The L1 loss and its standard deviation are shown directly on the top-left corner of error heat-maps. We refer readers to the supplementary video for better visualization. A large-version generator works as a baseline design to test with different ablative conditions.

Figure \ref{fig:ab_input} and Table \ref{table_renderer} (top) show the importance of our two conditional input. Removing upper body motions leads to strong shakiness especially around the neck and shoulders. It is hard for the network to synthesize time-consistent upper body motions without this strong location condition. Removing the candidate image set also leads to significant performance decrease on backgrounds and facial contours, especially when the training videos includes camera motions. For example, Obama video in Figure \ref{fig:ab_input} consists of several different camera motions and confuses the model to learn the one-to-many mappings. The candidate image set works as a clue to tell the model what the scene is and guides the network synthesize the right and consistent background. Besides, it also relieve the network pressure to synthesize high-fidelity details since it has included details as input. 

Figure \ref{fig:ab_architecture} and Table \ref{table_renderer} (middle) evaluate the generator architecture design. We compare our model with a baseline model Unet \cite{isola2017image} and a larger model with 2 residual blocks each layer. We found significant performance degradation using Unet (see red boxes in Figure \ref{fig:ab_architecture}). They fail to synthesize clear teeth, ears and other facial details compared to other architectures. Increasing the res-block in each layer takes over 59\% more parameters (121.790\emph{M} vs 76.204\emph{M}) but obtains no significant improvements on image quality. Considering the balance of performance and efficiency, we use normal architecture (1 Res-block) as our default architecture.

Finally we evaluate the importance of training dataset size. In this experiment, we train the model using 0.5/1/3 minutes frames (see Figure \ref{fig:ab_datasize}). Larger training set generates better results since it covers more pose and expression variations (see Table \ref{table_renderer} (bottom)), and the best results are achieved using the full training set.

\begin{figure*}[h]
    \centering
  \includegraphics[scale=0.41]{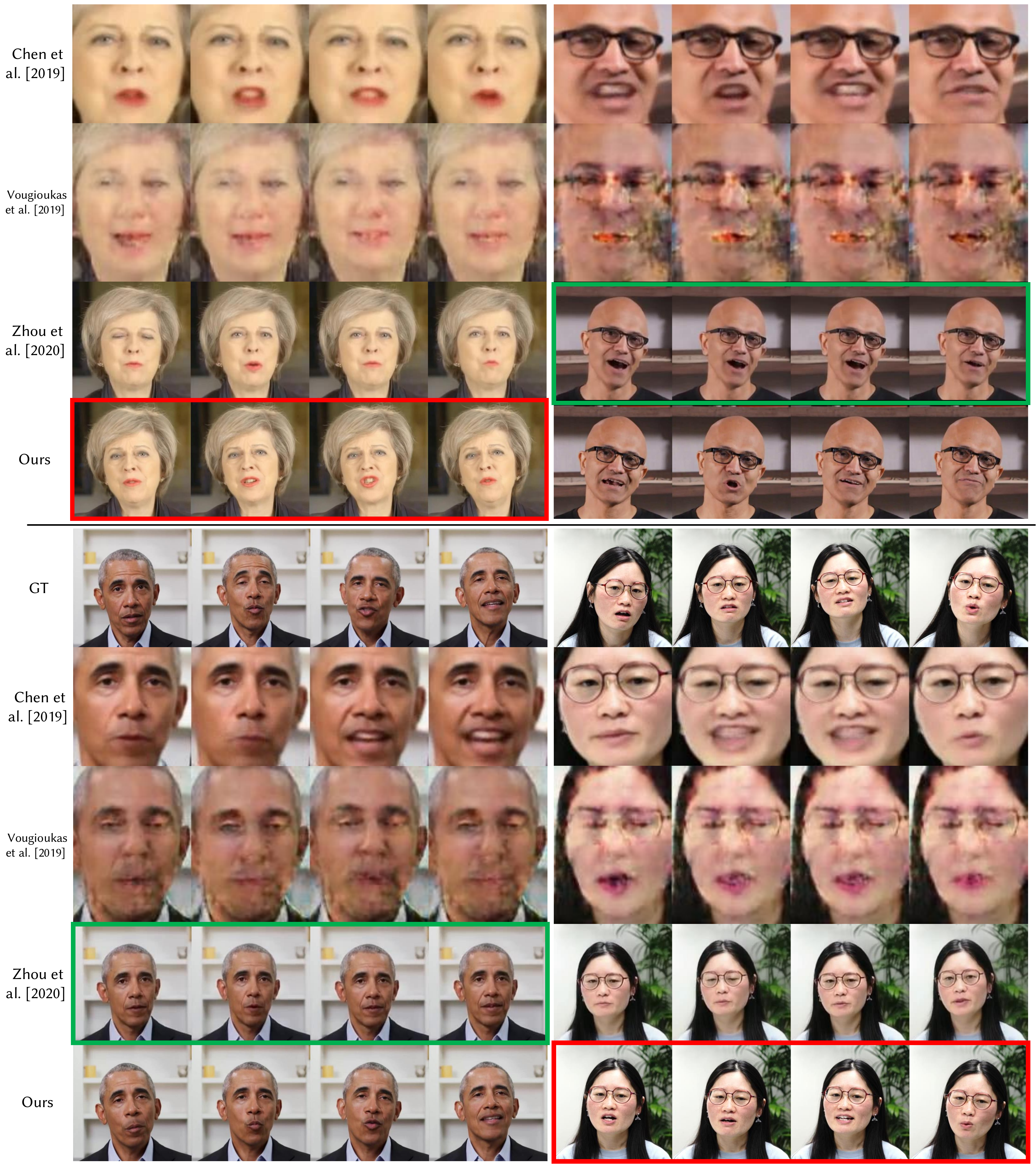}
  \caption{Comparison with state-of-the-art image-based generation methods. Best viewed in the supplementary video. Upper the solid line: test using wild audio streams. Lower the solid line: test using groundtruth audio streams in the validation set.
  Video \emph{May} \copyright \emph{UK government} (Open Government Licence). Video \emph{Nadella} \copyright \emph{IEEE Computer Society} (public domain). Video \emph{Obama} \copyright \emph{Barack Obama Foundation} (public domain).}
  \label{fig:compare_image}
\end{figure*}

\subsection{Comparisons to the State-of-the-Art}
Now we compare our method with state-of-the-art audio-driven talking-head animation techniques. All the test input audio sequences are unseen during training. We strongly recommend readers watch the supplementary video for better comparisons.\\

\emph{Comparisons to image-based generation methods.} We first compare our approach with image-based generation methods for synthesizing talking-head videos. Specially, we compare with Chen et al. \shortcite{ATVG}, Vougioukas et al. \shortcite{vougioukas2019realistic} and Zhou et al.  \shortcite{makelttalk}. Figure \ref{fig:compare_image} (upper the solid line) shows results driven by wild audios, and Figure \ref{fig:compare_image} (
lower the solid line) shows the results driven by voices of the target person in images. These methods are trained to generalize to unseen faces and therefore lack personal talking styles (they tend to generate the same lip motions for everyone) and facial details. Chen et al. \shortcite{ATVG} and Vougioukas et al. \shortcite{vougioukas2019realistic} generate the talking videos in a cropped and normalized face region and therefore fail to handle head poses. Zhou et al. \shortcite{makelttalk} generate speaker-aware talking-head videos but not the target person style. They warp both the background and the talking-head, giving the impression that the foreground head carries the background and moves together (see the green boxes). Moreover, the mouth of the portrait tends to twist and the synthesized region is blurred. Compared against these methods, our method successfully captures the talking style of the target personal and synthesizes sharper images with higher fidelity, e.g., wrinkles and teeth  (see the red boxes). We also note that the head pose generated by our method is different from the groundtruth but exhibits a similar distribution in the training set. 

We further compare with Chen et al. \shortcite{chen2020talking} in the supplementary video, which generate talking faces with rhythmic head movements. We notice that Chen et al. \shortcite{chen2020talking} do not disentangle the head motion and the background, resulting in an associated movement of the both. Besides, our method keeps the facial details and generates facial images with higher quality.\\

\emph{Comparisons to video-based editing methods.} We compare our approach against video-based editing methods. In particular, we compare against Suwajanakorn et al. \shortcite{SIG17Obama} and Thies et al. \shortcite{NVP}. Please watch the supplementary video for the results. These two methods both synthesize a lower face patch and blend it into the target frame. They rely on additional facial tracking algorithm to provide accurate and stable rigid head motions and 3D mouth locations. Our method directly generates full head renderings as well as backgrounds. Suwajanakorn et al. \shortcite{SIG17Obama} synthesize high-quality talking videos of Obama, which is trained on a huge amount of his weekly address videos (17 hours). \emph{Neural Voice Puppetry} \cite{NVP} additionally trains person-specific blendshape basis using 2-3 minute videos aside from training a general model basis on around 3 hours videos. Our method generates visually comparable and controllable results while using only around 3 minutes videos for training. Considering the intrinsic features of video-based editing methods, which limits the application scenarios, our approach is more applicable to other target person. Last but not least, our system runs in real-time.\\

\emph{Comparisons to model-based methods.} We also compare our approach with model-based methods \cite{SIG17NVIDIA, SIG17CMU, visemenet, VOCA}. These methods focus on learning a 3D face mesh or rigging parameters from audio. They usually require a 4D training corpus obtained by a high-cost vision-based capture system or rigging parameters with artist interventions, while our method uses sparse 3D landmarks as an intermediate representation and works on internet videos. Also, our method generates photorealistic results.

\begin{figure}[t]
  \centering
  \includegraphics[scale=0.8]{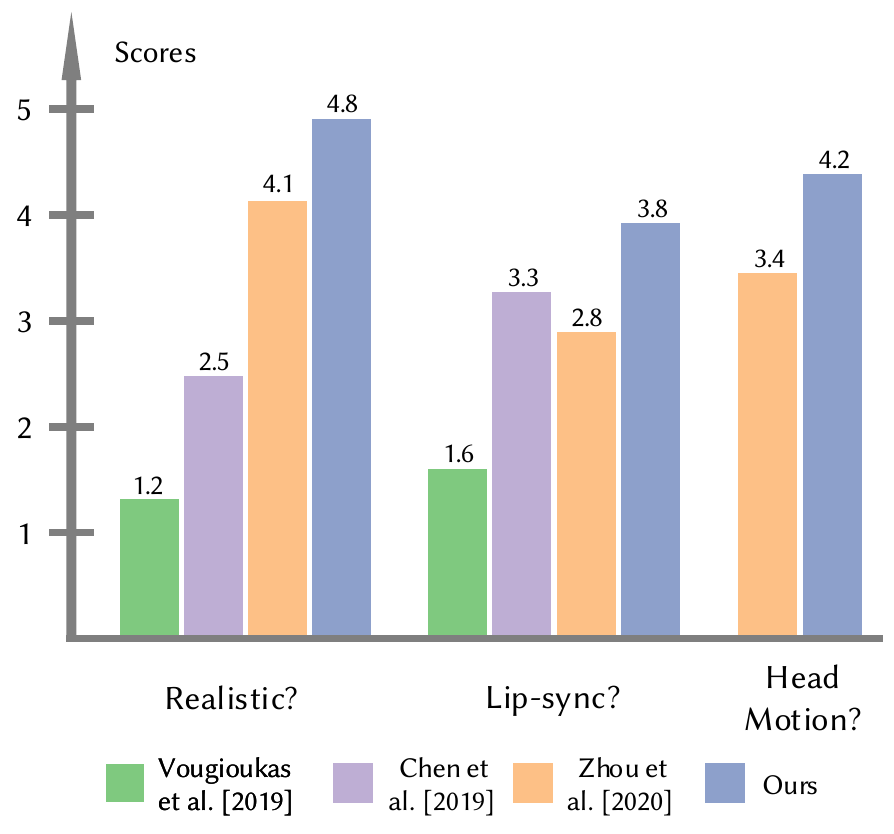}
  \caption{User study results for three different tasks.}
  \label{fig:user_study}
\end{figure}

\subsection{User Study}
We finally conduct three user studies to quantitatively evaluate the quality of our method. We compare our results with state-of-the-art open-source methods \cite{vougioukas2019realistic, ATVG, makelttalk}. We prepare 20 audio clips for each method and generate 80 video clips in total. All audio clips are wild and unseen in the training set. These user studies are web-based, and 48 participants with computer science backgrounds finished our questionnaire. During the study, the web page shows one video at a time in randomized order, and the participant is asked to evaluate the video w.r.t three statements: 'The video looks realistic to me.', 'The mouth motion is sync with the audio.' and 'The head motion of the portrait is natural.' on a scale between 1 to 5 (5-strongly agree, 4-agree, 3-neither agree nor disagree, 2-disagree, 1-strongly disagree). Figure \ref{fig:user_study} shows the average scores of different methods on three statements. The head pose evaluation is only conducted on Zhou et al. \shortcite{makelttalk} since the other two methods animate the cropped face and hardly have any pose motions.

\begin{figure}[t]
  \centering
  \includegraphics[scale=2.0]{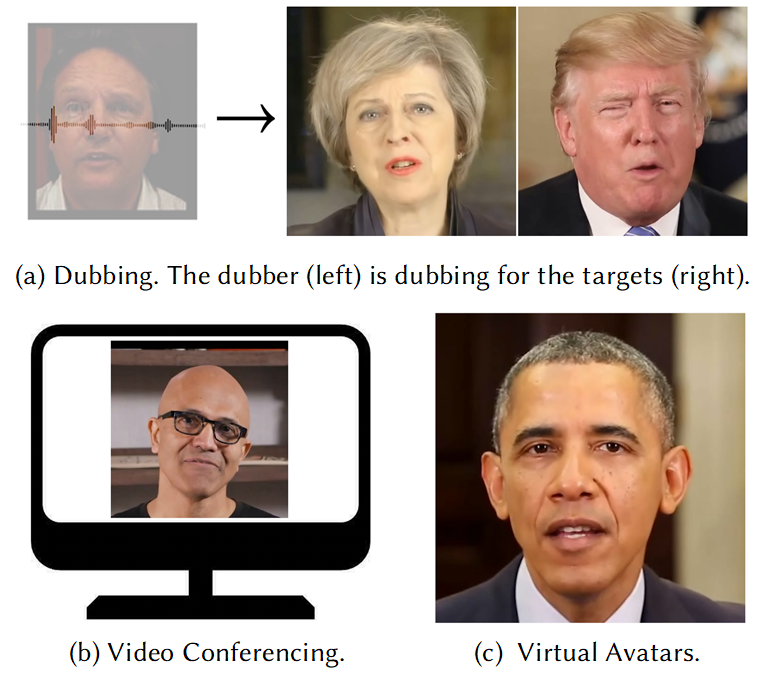}
  \caption{Applications. Our system can be applied on many applications, e.g., dubbing, video conferencing and virtual avatars. Please refer to the supplementary video for full sequences. 
  Video \emph{May} \copyright \emph{UK government} (Open Government Licence). Video \emph{Trump} \copyright \emph{White House} (public domain). Video \emph{Nadella} \copyright \emph{IEEE Computer Society} (public domain). Video \emph{Obama} \copyright \emph{White House} (public domain).}
  \label{fig:applications}
\end{figure}

As can be seen, our method achieves the highest scores over three tasks. For the first task, our method has the highest score of 4.8, which means our results are the most photorealistic. Meanwhile, we have the best results in terms of lip-synchronization. We believe that is because our manifold projection works which improves the generalization. Finally, we compare the head motion against Zhou et al. \shortcite{makelttalk}. Our method generates more natural head motions because we model the target-aware head motion.

\subsection{Applications}
Our method synthesizes photorealistic talking-head animation from audio streams in real-time, thus having a wide range of applications, e.g., dubbing, video conferencing, and virtual avatars. We refer readers to our supplementary video. Figure \ref{fig:applications} demonstrates the potential applications. On the top of the figure, we show the audio-driven dubbing results of the target person. Compared to video-based dubbing methods \cite{kim2019neural}, our method avoids generating implausible facial dynamics of the target person because we model the personal characteristics.

Video conferencing is another application (see Figure \ref{fig:applications} (b)). In scenarios that people cannot deliver visual signals, e.g., they are outdoor or have limited bandwidth, our method can generate high-fidelity video frames only driven by audio in real-time.

We finally demonstrate our potentials in virtual avatars such as virtual anchors, assistants. Our supplementary video shows a real-time demo of virtual avatars, e.g., the portrait of Theresa May is animated to sing a song driven by the actor's voice. Figure \ref{fig:applications} (c) shows results driven by the Text-to-Speech (TTS) system. The supplementary video also includes a comparison with Zhou et al. \shortcite{makelttalk} and Thies et al. \shortcite{NVP}. Our method generates more realistic frames and more accurate lip synchronization.

\section{Conclusion}
We presented a deep learning approach for generating photorealistic talking-head animation of the target person in real-time. Our method can handle new audio clips not seen during training and still synthesize personalized video frames. The full system is only needed to be trained on a several-minute length video. Our pipeline contains three stages: deep audio features extraction, facial dynamics and motion generation, and photorealistic image synthesis. The first stage includes a manifold projection on deep audio features, which helps generalize to wild audio. In the second stage, facial dynamics, head poses, and upper body motions are generated. An autoregressive probabilistic head pose estimation network is trained to learn the target actor's pose distribution. This network led to personalized head pose generation and avoided the potential performance degradation of the subsequent neural renderer. Finally, we generated intermediate feature maps from these predictions and sent them with a candidate image set to an image-to-image translation network to synthesize video frames. Thorough experiments and a user study show that our method outperforms the state-of-the-art techniques both qualitatively and quantitatively. Our method can be applied in many scenarios, especially the ones required to run in real-time, like dubbing, video conferencing, and virtual avatars. We hope this work could open a new avenue for future researches in this field.\\

\emph{Limitations and Future Work.} While we have demonstrated impressive results of our method in a wide variety of scenarios, there are still several limitations to our approach. Our real-time system does not always capture well with the plosive and nasal consonants, e.g., /p/b/m/. The reasons behind are various. First, /p/b/m/ usually sounds in low volumes and may be ignored by the fore-end as environmental noise. Second, our live system runs at over 30 FPS, and it may miss these short sounds. It also fails to capture the speakings at a very fast speed, like a quarrel scenario. Our offine results (60FPS) are better, which partly verifies our supposition. Applying model pruning is a promising solution to decrease the parameters and increase the running speed. Besides, the spectrum construction we use tends to miss those short phonemes, which can be tackled by using pure deep features, e.g., wav2vec \cite{wav2vec}. The face tracking algorithm we used is not state-of-the-art, we
believe that better reconstruction leads to better lip-sync results.


Similar to most learning-based methods, the style of the generated videos is restricted to the training corpus. Our method preserves the talking styles in training sequences (3-5 minutes) via manifold projection, a domain transfer method to find the most similar samples. This mechanism alleviates this problem to some extent. We believe that a complete solution is to apply a perfect audio disentanglement algorithm like \cite{qian2020unsupervised} to split each components, i.e., content, pitch, timbre and rhythm, and find the best mapping of these components.

Emotional audio may generate unsatisfied results when the model was trained on a neutral-style video. Our method cannot directly control the emotion of the generated videos. Recent work \cite{ji2021audio} shows promising emotion manipulation results when training on an emotional dataset. It would be interesting to apply such progress to our system.

Although we successfully handle the shadows and the lighting reflections when people swing their heads, we still can not explicitly control these parameters. Relighting techniques \cite{sun2019single} can be applied directly to our rendering results to control the environment lighting. Gestures are another important component for people to deliver expression. We are looking forward to future work on gesture generation driven by audio.

\section{Ethical Considerations}
With the rapid development of 'Deepfake' techniques, the threshold for people to synthesize fake videos of arbitrary person is becoming lower. In most kinds, they facilitate the movie and entertainment industry and reduces the bandwidth of video streaming by sending the audio signals only. However, these techniques can be misused. Due to the fact that it's more difficult for people to distinguish fake videos, the algorithm may be utilized to spread misinformation or obtain illegal profits. Our method achieves real-time photorealistic talking-head animation and only needs to be trained on a several-minute length video, which can be easily found on the Internet. For non-celebrities, their faces and voices are harder to recognize than celebrities, and therefore generating their fake videos is more deceptive. Potential solutions like digital face forensics methods \cite{rossler2018faceforensics, rossler2019faceforensics++} to detect deepfakes must be considered. We hope the public be aware of the potential risks of the misuse of new techniques.

\section*{Acknowledgments}
\addcontentsline{toc}{section}{Acknowledgments}
We would like to thank Shuaizhen Jing for the help with the Tensorrt implementation. We are grateful to Qingqing Tian for the facial capture. Yuanxun Lu would also like to thank Xinya Ji for her mental support and proof-reading during the project. This work was supported by the NSFC grant 62025108, 61627804 and Leading Technology of Jiangsu Basic Research Plan (BK20192003).

\bibliographystyle{ACM-Reference-Format}
\bibliography{bibliography}

\appendix
\section{APPENDIX}
\label{appendix}

In this appendix, we describe all the sequences we used in detail (Table \ref{table_dataset}).

\begin{table}[ht]
    \caption{List of dataset used in our experiments. Video \emph{May} \copyright \emph{UK government} (Open Government Licence). Video \emph{Obama1} \copyright \emph{White House} (public domain). Video \emph{Obama2} \copyright \emph{Barack Obama Foundation} (public domain). Video \emph{Nadella} \copyright \emph{IEEE Computer Society} (public domain). Video \emph{Trump} \copyright \emph{White House} (public domain). Video \emph{Ford} \copyright \emph{Ontario Office} (public domain). Video \emph{McStay} \copyright \emph{Darren McStay} (CC BY).}
    \label{table_dataset}
    \begin{tabular}{c c c}
    Video & Name & Length \\ \hline
    \begin{minipage}[b]{0.1\columnwidth}
		\centering
		\raisebox{-.5\height}{\includegraphics[width=\linewidth]{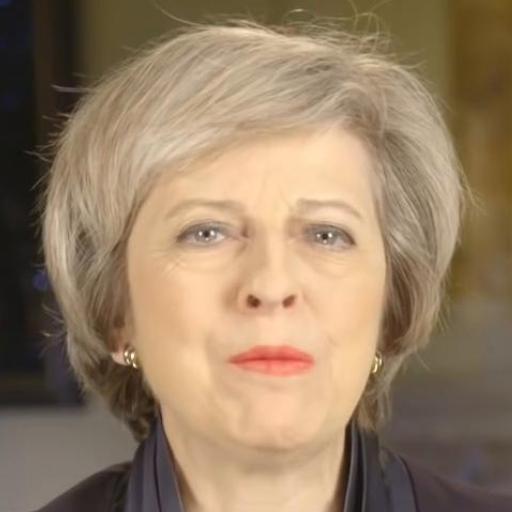}}
	\end{minipage}
	& May & 4min 02s \\
    \begin{minipage}[b]{0.1\columnwidth}
		\centering
		\raisebox{-.5\height}{\includegraphics[width=\linewidth]{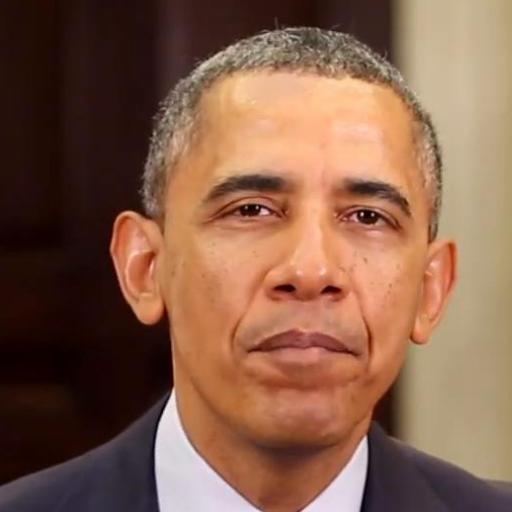}}
	\end{minipage}
	& Obama1 & 2min 59s \\
    \begin{minipage}[b]{0.1\columnwidth}
		\centering
		\raisebox{-.5\height}{\includegraphics[width=\linewidth]{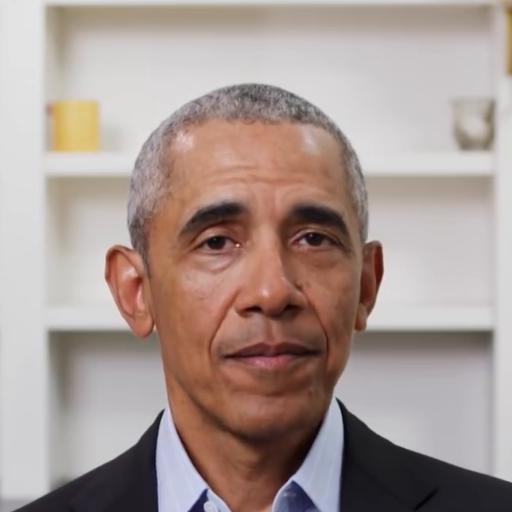}}
	\end{minipage}
	& Obama2 & 3min 42s \\
    \begin{minipage}[b]{0.1\columnwidth}
		\centering
		\raisebox{-.5\height}{\includegraphics[width=\linewidth]{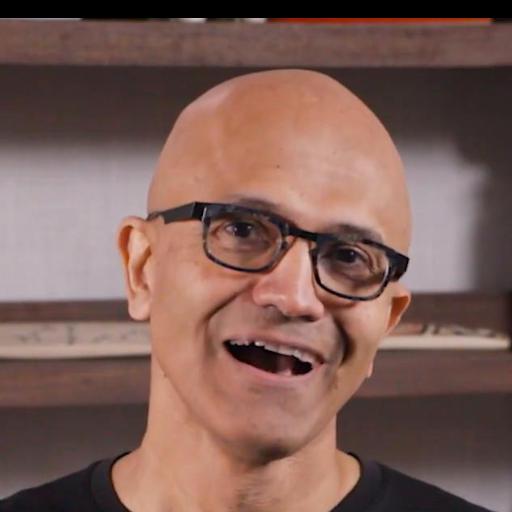}}
	\end{minipage}
	& Nadella & 3min 9s \\
    \begin{minipage}[b]{0.1\columnwidth}
		\centering
		\raisebox{-.5\height}{\includegraphics[width=\linewidth]{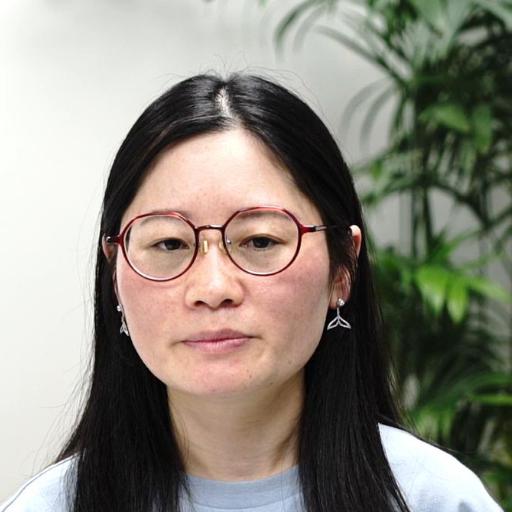}}
	\end{minipage}
	& Actor A & 3 min 45s \\
    \begin{minipage}[b]{0.1\columnwidth}
		\centering
		\raisebox{-.5\height}{\includegraphics[width=\linewidth]{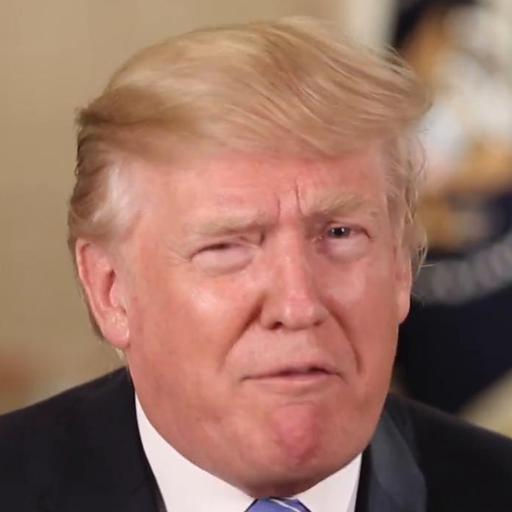}}
	\end{minipage}
	& Trump & 3min 31s \\
    \begin{minipage}[b]{0.1\columnwidth}
		\centering
		\raisebox{-.5\height}{\includegraphics[width=\linewidth]{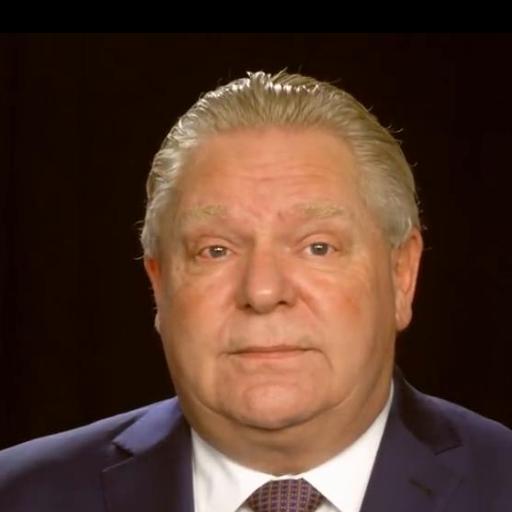}}
	\end{minipage}
	& Ford & 3 min 10s \\
    \begin{minipage}[b]{0.1\columnwidth}
		\centering
		\raisebox{-.5\height}{\includegraphics[width=\linewidth]{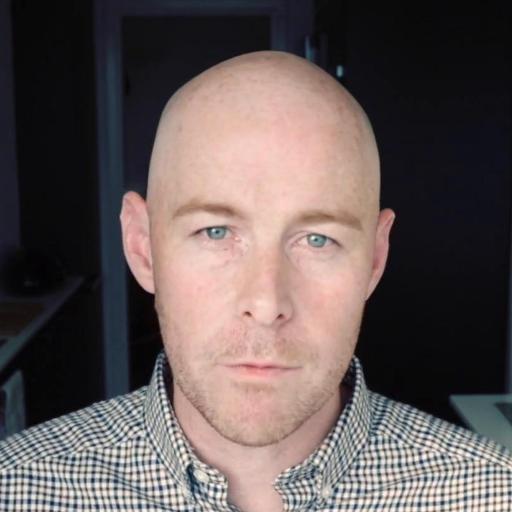}}
	\end{minipage}
	& McStay & 4min 30s \\
    \end{tabular}
\end{table}